\documentclass[aps,pre,preprint,eqsecnum,amssymb,floatfix]{revtex4}
\usepackage{graphicx,color,psfrag}
\usepackage{amsmath}
\usepackage{bm}
\begin{document}
\title{On the Redfield and Lindblad master equations}
\author{Hans C. Fogedby}
\email{fogedby@phys.au.dk,  hans.fogedby@gmail.com}
\affiliation{Department of Physics and Astronomy, 
University of Aarhus, Ny Munkegade,
8000 Aarhus C, Denmark}
\begin{abstract}
In a previous work we developed a field theoretical approach to open
quantum systems using condensed matter methods. In the Born
approximation we derived the Redfield equation on the basis of a 
multi-oscillator bath, a Dyson equation, a diagrammatic expansion 
and a quasi-particle approximation. In addition, applying a rotating 
wave approximation, we obtained the Lindblad equation describing
a proper quantum map. The issue regarding the additional 
rotating wave approximation was left as an open problem.

The present work addresses the open problem and presents new results. 
We identify a discrepancy in the popular and standard Redfield equation. 
The discrepancy is associated with the well-known fact that the Redfield 
equation does not represent a proper quantum map.
The discrepancy is related to the diagrammatic expansion and a 
consistency requirement in the 
quasi-particle approximation. The explicit resolution of this discrepancy
is obtained by imposing energy conservation on the Born level.
As a result we obtain formal equivalence between the
energy-conserving Redfield equation and the Lindblad
equation without invoking the rotating wave approximation. 
We provide a detailed mapping of the field theoretical
approach to the standard microscopic derivation in the theory
of open quantum systems.
\end{abstract}
\maketitle
\section{\label{intro}Introduction}
There is a current interest in open quantum systems that is quantum systems
interacting with a prescribed environment or bath. There is in particular a large 
body of recent literature in the context of quantum entanglement, see e.g. 
\cite{Horodecki09,Paneru20,Bresque21} and quantum thermodynamics, see e.g. 
\cite{Vinjanampathy16,Kosloff19,Rivas20}.

In the present work we will focus on the two primary analytical tools
used in the theory of open quantum systems, namely the Redfield and
Lindblad master equations. Both equations are based on the assumption of
a weak interaction between the quantum system and the bath and the assumption 
of a time scale separation between system and bath. The equations
both operate in the memoryless Markov limit.

For details we refer in the following to the textbook by Breuer and Petruccione 
\cite{Breuer06}. In the customary approach, the Redfield equation 
\cite{Redfield65} is easily derived in the Born approximation on the basis of the  
von Neumann master equation, combined with a series of physical approximations. 
A more systematic approach is based on the Nakajima-Zwanzig projection 
operator techniques, see e.g. \cite{Gonzalez24}. The Redfield equation can
also be obtained from the perturbative time-convulation-less method,
see e.g.  \cite{Breuer01}.

The Redfield equation preserves the trace of the density operator for
the quantum systems but does no guarantee positivity and thus does
not correspond to a proper quantum map. However, by applying
the rotating wave approximation (RWA) one obtains the Lindblad equation
which from general principles describes a quantum map 
\cite{Lindblad76,Manzano20,Chrus17,Tupkary22};
the Lindblad equation is also often referred to as the quantum optical
master equation, the GKSL equation 
 \cite{Gorini76,Kossakowski21,Sudarshan63,Lindblad76} or the Davies equation
 \cite{Davies74}.
 The focus of the present work is to reanalyse the connection
 between the Redfield and Lindblad equations and in particular
 the RWA using field theoretical methods. 
\subsubsection{Previous work}
In previous work \cite{Fogedby22} we developed a field theoretical approach 
based on condensed matter methods.
Assuming that the  bath is composed of independent quantum oscillators,
i.e., the Caldeira-Leggett prescription \cite{Caldeira83a},
we derived a general master equation for the reduced density operator 
$\rho_S(t)$,
\begin{eqnarray}
\frac{d\rho_S(t)}{dt}=
-i[H_S,\rho_S(t)]+\int^{\infty}_{-\infty} dt' K(t,t')\rho_S(t'),
\label{firstmaster}
\end{eqnarray}
where  $H_S$ is the Hamiltonian for the open quantum system;
note that in the above equation and in the subsequent expressions 
we set the Planck constant $\hbar$ equal to unity ($\hbar=1$).

The kernel $K(t,t')$ encapsulates all information regarding the 
coupling to the bath such as dissipation and energy shifts.
Based on a diagrammatic representation the kernel is determined 
by an expansion in powers of the interaction between the system and 
the bath. 
In order to implement the Markov limit, we applied  the condensed 
matter quasi-particle approximation locking $K(t,t')$ in Fourier
space, $\tilde K(\omega)$, onto the energy spacings, $E_n$,
of the open quantum system, i.e. 
$\tilde K(\omega)\to \tilde K(E_n)=\tilde K$.
This mapping corresponds to  setting $K(t,t')=\tilde K\delta(t-t')$, yielding the 
Markov master equation
\begin{eqnarray}
\frac{d\rho_S(t)}{dt}=
-i[H_S,\rho_S(t)]+\tilde K\rho_S(t).
\label{secondmaster}
\end{eqnarray}
In the Born approximation to second order in the system-bath
interaction, the kernel $K(t,t')$ is composed of four contributions involving
a single bath correlation function. Implementing the quasi-particle
approximation, we obtain the Redfield equation. However, in order 
to obtain the Lindblad equation the rotating wave approximation 
(RWA) is invoked,  as is also done in the standard derivation
\cite{Breuer06}. 

This ad hoc application of the RWA takes us outside 
the field theoretical  framework and was left as an open issue.
In the present paper we address this issue.
\subsubsection{Present work}
The RWA issue was left unresolved in our previous work. In the present
work we present new results not previously established in the literature.
We thus identify a discrepancy in the standard derivation of the Redfield 
equation. Using an ambiguity in the quasi-particle approximation this 
discrepancy is subsequently resolved by imposing "energy conservation 
on the Born level", corresponding to a selection rule applied to the individual
terms in the perturbation expansion.
Accounting for this
discrepancy then readily yields the Lindblad equation without invoking
the RWA. This is our main result.
\subsubsection{Layout}
For the benefit of the reader we summarise in Sec. \ref{review} aspects of 
the field theoretical 
approach. In Sec. \ref{new} we present and discuss the new results. 
In Sec. \ref{discussion} we present a summary, discuss the relation
to quantum field theory, consider 
populations and  coherences, compare the present approach with
the equation of motion method (EQM), the Nakajima-Zwanzig approach
(NZ), the Time-convolutionless projection method (TCL) and  finally 
comment on  related work, in particular the recently proposed 
{\em universal Lindblad  equation}.  In Appendix \ref{appendix} we discuss in more 
detail  i) the transmission matrix, ii) the EQM, iii) the NZ, and iv) the TCL.
\section{\label{review}Previous work}
In this section we summarise components of the field theoretical approach 
to be used in the present analysis. Details are available in the field theoretical 
treatment in \cite{Fogedby22}, see also  Appendix \ref{appendix}.
\subsection{Open quantum system}
An open quantum system is described by the global Hamiltonian
\begin{eqnarray}
H=H_S+H_B+H_{SB},
\label{hamiltonian}
\end{eqnarray}
where $H_S$ is the system Hamiltonian, $H_B$ the 
bath Hamiltonian, and $H_{SB}$ the system-bath interaction Hamiltonian.
The system and bath operators are denoted by $S^\alpha$ and $B^\alpha$, 
respectively. From the Hilbert spaces associated with the bath and open 
quantum system the Schmidt Decomposition theorem applied to $S^\alpha$ 
and $B^\alpha$ implies a system-bath coupling of the form
$H_{SB}=\sum_\alpha S^\alpha B^\alpha$ \cite{Zhang24}. In a physics 
context this corresponds to a dipole-type system-bath coupling.
Moreover, in order to render the analysis precise we proceed in the following
in the energy basis of the system 
Hamiltonian $H_S$, i.e., $H_S|n\rangle=E_n|n\rangle$, $E_{pq}=E_p-E_q$;
here $|n\rangle$ are the energy eigenstates, $E_n$ the associated energy 
levels and $E_{pq}$ the energy splittings.
\subsection{Transmission operator}
Focussing on the dynamical map $T$, the evolution of the density operator for 
the quantum system $\rho_S(t)_{pp'}$ from an initial state at time $t_i$ to 
a final state at time $t$, is 
characterised by the transmission matrix $T(t,t_i)_{pp',qq'}$ according to 
\begin{eqnarray}
\rho_S(t)_{pp'}=\sum_{qq'}T(t,t_i)_{pp',qq'}\rho_S(t_i)_{qq'}.
\label{density}
\end{eqnarray}
Above and in the following we make use of the shorthand notation
$\rho_S(t)_{pp'}=\langle p|\rho_S(t)|p'\rangle$ and 
$T(t,t')_{pp',qq'}=\langle p|\langle q'| T(t,t')|q\rangle|p'\rangle$;
note that $T(t,t')$ is a super operator acting on the operator $\rho_S$

We assume that the bath is static and described by the density operator
$\rho_B$. As a consequence the system exhibits time translational
invariance. However, for convenience we make use of the compact
notation $T(t,t')\equiv T(t-t')$ etc.

The transmission operator in (\ref{density}) is equivalent to the general Kraus 
representation $K_{BB'}(t)$ based on a spectral representation of the density
operator $\rho_B$ for  the bath  \cite{Breuer06}.  Assuming 
$\rho(0)=\rho_S(0)\rho_B$, where the total density operator $\rho(t)$ evolves
is time according to $\rho(t)=U(t)\rho(0)U^\dagger(t)$, $U(t)=\exp(-iHt)$, and 
inserting the spectral representation 
$\rho_B=\sum_{B'}E_{B'}|B'\rangle\langle B'|$, we have 
$\rho_S(t)=\sum_{BB'}K_{BB'}(t)\rho_S(0)K^\dagger_{BB'}(t)$, where
$K_{BB'}(t)=\sqrt{E_{B'}}\langle B|U(t)|B'\rangle$. In the energy representation
the Kraus representation yields the transmission matrix in (\ref{density}).
\subsection{Oscillator bath, Wicks theorem and Dyson's equation}
The next step is to assume that the bath is composed of 
a set of quantum oscillators, 
i.e., the Caldeira-Leggett prescription
\cite{Caldeira83a}, implying the validity of Wick's theorem \cite{Gaudin60}.
Wick's theorem implies that multi-bath operator correlations break
up into products of two-bath operator correlations.
Consequently, $T(t,t')_{pp',qq'}$ satisfies a fundamental Dyson equation 
incorporating secular effects
\begin{eqnarray}
T(t,t')_{pp',qq'}=
T^0(t,t')_{pp',qq'}+
\sum_{ss',ll'}\int^{\infty}_{-\infty} dt''dt''' T^0(t,t'')_{pp',ss'}
K(t'',t''')_{ss',ll'}T(t''',t')_{ll',qq'}.~~
\label{dyson}
\end{eqnarray}
The free transmission matrix $T^0(t,t')_{pp',qq'}$ is given by
\begin{eqnarray}
&&T^0(t,t')_{pp',qq'}=G_R(t,t')_{pq}G_A(t',t)_{q'p'}.
\label{free}
\end{eqnarray}
We have introduced the retarded and advanced Green's functions
\begin{eqnarray}
&&G_R(t,t')_{pq}=-i\eta(t-t')\delta_{pq}\exp(-iE_p(t-t')),
\label{greenret}
\\
&&G_A(t,t')_{pq}=+i\eta(t'-t)\delta_{pq}\exp(-iE_p(t-t')),
\label{greenadv}
\end{eqnarray}
where $\eta(t)$ is the step function: $\eta(t)=1$ for $t>0$, $\eta(t)=0$ for $t<0$,
$\eta(0)=1/2$, and $d\eta(t)/dt=\delta(t)$. 
The transmission matrix $T^0(t,t')_{pp',qq'}$ describes the evolution of
$\rho_S(t)$ in the absence of coupling to the bath.
\subsection{Master equation and kernel}
It next follows that the Dyson equation (\ref{dyson}) implies an inhomogeneous 
integral equation for $\rho_S(t)_{pp'}$,
depending on the initial assigment $\rho_S(t_i)_{qq'}$,
\begin{eqnarray}
\rho_S(t)_{pp'}=\sum_{qq'}T^0(t,t_i)_{pp',qq'}\rho_S(t_i)_{qq'}+
\sum_{ss',qq'}\int^{\infty}_{-\infty} dt'dt'' 
T^0(t,t')_{pp',ss'}K(t',t'')_{ss'qq'}\rho_S(t'')_{qq'},
\nonumber
\\
\label{intmaster}
\end{eqnarray}
and since 
\begin{eqnarray}
\frac{dT^0(t,t')_{pp',qq'}}{dt}=-iE_{pp'}T^0(t,t')_{pp',qq'}+\
\delta(t-t')\delta_{pq}\delta_{p'q'},
\label{free2}
\end{eqnarray}
 a general master equation for 
$\rho_S(t)_{pp'}$,
\begin{eqnarray}
\frac{d\rho_S(t)_{pp'}}{dt}=
-iE_{pp'}\rho_S(t)_{pp'}+
\sum_{qq'}\int^{\infty}_{-\infty} dt' K(t,t')_{pp',qq'}\rho_S(t')_{qq'};
\label{master}
\end{eqnarray}
note that in deriving (\ref{master}) we have discarded the term 
$\delta(t-t_i)\rho_S(t_i)_{pp'}$.  This is justified  since $K$  determined
by the bath correlations has a finite
range $\tau$ and we, moreover, assume that the master equation operates 
in a range given by $t\gg t_i$. 

For later purposes, we also need the algebraic relations following from
a Fourier transformation, i.e., $\tilde T(\omega)_{pp',qq'}=
\int^{\infty}_{-\infty} d\tau\exp(i\omega\tau)T(\tau)_{pp',qq'}$, 
$\tau=t-t'$, etc. We have
\begin{eqnarray}
&&\tilde T(\omega)_{pp',qq'}=\tilde T^0(\omega)_{pp',qq'}+
\sum_{ss',ll'}\tilde T^0(\omega)_{pp',ss'}
\tilde K(\omega)_{ss',ll'}\tilde T(\omega)_{ll',qq'},
\label{four-dyson}
\\
&&-i\omega\tilde\rho_S(\omega)_{pp'}=  
-iE_{pp'}\tilde\rho_S(\omega)_{pp'}+
\sum_{qq'}\tilde K(\omega)_{pp',qq'}\tilde\rho_S(\omega)_{qq'}.
\label{four-master}
\end{eqnarray}
The master equation (\ref{master}) is a fundamental result. It follows solely
from a multi-oscillator heat bath. The first term describes the unitary evolution
of the density matrix in the absence of  coupling to the bath, the second term
contains the kernel which characterises the coupling to the bath. The kernel
is in general complex and gives rise to both dissipation and an energy shift.

The expressions above are depicted in Fig. 1. The relationship in
(\ref{density}) shown in Fig. 1a, the Green's functions (\ref{greenret}) and
(\ref{greenadv}) as directed arrows in Fig. 1b, the Dyson equation (\ref{dyson}) 
in Fig. 1c  and the integral equation (\ref{intmaster}) in Fig. 1d.
\subsection{Diagrammatic perturbation theory - Born approximation}
The diagrammatic representation in Fig. 1 allows for an expansion of the 
kernel $K$ in powers of the system-bath interaction $H_{SB}$. The kernel 
is composed of a real component giving rise to dissipation and an imaginary 
components which can be incorporated in the Hamiltonian $H_S$ as a 
Lamb-type energy shift; in the following we consider only the dissipative part. 

In the Born approximation depicted in Fig. 1e we obtain four contributions 
to the dissipative component:
\begin{eqnarray}
\tilde K(\omega)_{pp',qq'}=
&&-\frac{1}{2}\delta_{p'q'}
\sum_{\alpha\beta,l}S^\alpha_{pl}S^\beta_{lq}
\tilde D^{\alpha\beta}(\omega+E_{q'l})
\nonumber
\\
&&-\frac{1}{2}\delta_{pq}
\sum_{\alpha\beta,l}S^\alpha_{q'l}S^\beta_{lp'}
\tilde D^{\alpha\beta}(-\omega+E_{ql})
\nonumber
\\
&&+\frac{1}{2}
\sum_{\alpha\beta}S^\beta_{pq}S^\alpha_{q'p'}
\tilde D^{\alpha\beta}(-\omega+E_{qp'})
\nonumber
\\
&&+\frac{1}{2}
\sum_{\alpha\beta}S^\beta_{pq}S^\alpha_{q'p'}
\tilde D^{\alpha\beta}(\omega+E_{q'p}).
\label{kernelborn}
\end{eqnarray}
Here the bath correlation function and its Fourier transform are given by
\begin{eqnarray}
&&D^{\alpha\beta}(t,t')=\text{Tr}[\rho_BB^\alpha(t)B^\beta(t')],
\label{bath}
\\
&&\tilde D^{\alpha\beta}(\omega)=\int^{\infty}_{-\infty} d\tau\exp(i\omega\tau)
D^{\alpha\beta}(\tau), \tau=t-t',
\label{fourier-bath}
\end{eqnarray}
where $\rho_B$ is the density operator for the bath; we note that the static 
bath implies time translation invariance. We shall also later make use 
of the notation $\tilde D^{\alpha\beta}(\omega)\equiv
\langle B^\alpha B^\beta\rangle(\omega)$.  

Referring for more details to \cite{Fogedby22}, the bath Hamiltonian for a single
bath is given by $H_B=\sum_k\omega_k b_k^\dagger b_k$, where
$b_k$ is the Bose operator and $\omega_k$ the associated frequency.
The bath at temperature $1/\beta$ is characterised by the density 
operator $\rho_B=\exp(-\beta H_B)/Z_B(\beta)$,  
$Z_B(\beta)=\text{Tr}_B[\exp(-\beta H_B)]$. The bath operator 
$B(t)=\sum_k\lambda_kb_k(t)$, where the system-bath coupling
strength is incorporated in $\lambda_k$.

For the bath correlations in frequency space we obtain
\begin{eqnarray}
&&\langle B B^\dagger\rangle(\omega)=g(\omega)(1+n(\omega)), ~~~\omega>0,
\label{Bath1-Fourier}
\\
&&\langle B^\dagger B\rangle(\omega)=g(-\omega)n(-\omega),~~~ \omega<0,
\label{Bath2-Fourier}
\end{eqnarray}
where we have introduced the spectral strength $g(\omega)$ and the
Planck distribution $n(\omega)$,
\begin{eqnarray}
&&g(\omega)=2\pi\sum_k\lambda_k^2\delta(\omega-\omega_k),~~\omega>0,
\label{Spectral-Dis}
\\
&&n(\omega)=\frac{1}{\exp(\beta\omega)-1}, ~~\omega>0;
\label{Planck2}
\end{eqnarray}
the bath correlation function is depicted in Fig. 1b.

By inspection of (\ref{kernelborn}) we find that the trace condition
$\sum_p\tilde K(\omega)_{pp,qq'}=0$, following from 
$\sum_p\rho_S(t)_{pp}=1$, is satisfied. In terms of the numbering 
of the diagrams in the Fig. 1e the
trace condition follows by setting $p=p'$ in (1) and (4) and in (2) and (3)
and noting the cancellations.

We also note that in an open quantum system context regarding the
standard derivations \cite{Breuer06} the Born approximation also includes 
the decoupling assumption $\rho(t)\sim\rho_S(t)\rho_B$. 
This assumption
based on a weak coupling argument is here replaced by the Dyson
equation and the ensuing analysis.
\subsection{Quasi-particle approximation and the Markov limit}
The quasi-particle approximation was discussed in  \cite{Fogedby22}. Here
we expand our discussion. In condensed-matter many-body theory 
\cite{Mahan90} the basic building block is the single-particle Green's function 
$\tilde G(\omega,k)$ describing the propagation of a quantum particle with 
energy $\omega$ and momentum $k$ in a many-body environment. 
Suppressing the $k$ dependence the Green's function for a noninteracting system
has the form $\tilde G_0(\omega)=1/(\omega-E)$, where $E$ is an energy level. 
In the simplest case, diagrammatic perturbation theory gives rise to a Dyson equation 
of the form $\tilde G(\omega)=\tilde G_0(\omega)+
\tilde G_0(\omega)\tilde\Sigma(\omega)\tilde G(\omega)$,
whose solution is the generic form $\tilde G(\omega)=1/[\omega-E-\tilde\Sigma(\omega)]$;
here the self-energy or mass operator $\tilde\Sigma(\omega)$ is determined perturbatively.
The time dependence of the propagation of the quasi-particle is thus given by 
the resonance condition $\tilde G(\omega)^{-1}= \omega-E-\tilde\Sigma(\omega)=0$. 
To leading order, the quasi-particle approximation corresponds to $\tilde\Sigma(\omega) \to \tilde\Sigma(E)$. Separating $\tilde\Sigma(E)$ in a real and imaginary part, i.e., $\tilde\Sigma(E)=\tilde\Delta+i\tilde\Gamma$, we have
$G(t)\propto e^{-i(E+\tilde\Delta)t}e^{-\tilde\Gamma t}$; the real part $\tilde\Delta$ gives rise to a 
quasi-particle energy shift, while the imaginary part $\tilde\Gamma$ yields a damping of the quasi-particle. Both energy shift and damping are caused by interaction with the many-body environment. For the quasi-particle to preserve its identity, we must assume that the damping is small. It is important to note that the Dyson equation automatically incorporates secular effects in producing an energy shift and a damping.
In order to implement the Markov approximation yielding a kernel of 
the form $K(t,t')_{pp',qq'}\to K_{pp',qq'}\delta(t-t')$ or, equivalently, 
$\tilde K_{pp',qq'}(\omega)\to\tilde K_{pp',qq'}$ in (\ref{master}) and 
(\ref{four-master}), respectively, we turn to the quasi-particle approximation 
in condensed matter theory discussed above. 

Here we discuss the various time scales characterising an open quantum
system. Since we have set $\hbar=1$ energy is equivalent to inverse time
and we can discuss the various time scales in terms of the associated
frequencies. There are  three characteristic frequencies associated
with an open quantum system, $\omega_S$, $\omega_B$ and $\omega_{SB}$. 
The frequency characterising the quantum system is basically the energy 
splitting $E_{nm}$, i.e., $\omega_S\propto E_{nm}$. The system-bath coupling
is given by $\lambda_k$, i.e., $\omega_{SB}\propto \lambda_k$. The behaviour
of the bath reflected in the bath correlations is controlled by the spectral
function $g(\omega)$ and the Planck distribution $n(\omega)$.
For convergence reasons $g(\omega)$ must be cut-off at a characteristic
frequency $\omega_c$, e.g. $g(\omega)\propto\exp(-\omega/\omega_c)$,
i.e., $\omega_B\propto\omega_c$; note that the Planck distribution also
sets a frequency scale given by $T=1/\beta$, which set in for $T$ of order 
$\omega_B$.

Defining the inverse transmission matrix 
according to 
$\sum_{ll'}\tilde T_{pp',ll'}(\omega)^{-1}\tilde T_{ll',qq'}(\omega)=
\delta_{pq}\delta_{p'q'}$ we infer from (\ref{four-dyson})
\begin{eqnarray}
\tilde T(\omega)^{-1}_{pp',qq'}=
-i\delta_{pq}\delta_{p'q'}(\omega-E_{pp'})
-\tilde K(\omega)_{pp',qq'},
\label{inv-dyson}
\end{eqnarray}
where we from the unperturbed transmission matrix  in (\ref{free}) have 
inserted the expression
\begin{eqnarray}
T^0(\omega)^{-1}_{pp',qq'}=
-i\delta_{pq}\delta_{p'q'}(\omega-E_{pp'}).
\label{inv-free}
\end{eqnarray}
The time dependence of $T(t,t')_{pp',qq'}$ is thus determined by 
the resonance condition 
\begin{eqnarray}
\text{det}[-i\delta_{pq}\delta_{p'q'}(\omega-E_{pp'})-
\tilde K(\omega)_{pp',qq'}] =0,
\label{resonance}
\end{eqnarray}
yielding pole and branch cut contributions to $T(t,t')_{pp',qq'}$, 
i.e., the dynamical spectrum of the system.

Schematically, the dynamical spectrum is determined by the
condition $\omega-E-i\tilde K(\omega)=0$, and we infer $\omega_S=E$.
Assuming a weak system-bath coupling, $\omega_{SB}\ll \omega_S$,
we expand $K(\omega)$ about $E$ to second order, 
$K(\omega)=K(E)+K'(E)(\omega-E)$, corresponding to the
Born approximation. We thus obtain the resonance condition 
$(\omega-E)(1-iK'(E))-iK(E)=0$.
Next, assuming that $K(\omega)$ is flat about $E$ we set
$K'(E)=0$ and obtain the condition $\omega-E-iK(E)=0$.
The slow variation of $K(\omega)$ on the $E$ scale is obtained
by assuming $\omega_B\gg \omega_S$. The last condition is
equivalent to the time scale separation.

The time scale of the quantum system is basically given by $1/E_{pp'}$, 
whereas the bath time scale is determined by $\tilde K(\omega)_{pp',qq'}$. 
A fast time scale of the bath corresponds to a slow dependence on 
$\omega$; note that $K(t)_{pp',qq'}\propto\delta(t)$, yields 
$\tilde K(\omega)_{pp',qq'}\propto\text{const.}$. Assuming a time 
scale separation we can to leading order in $H_{SB}$ in (\ref{resonance})
replace the $\omega$ dependence in $\tilde K(\omega)_{pp',qq'}$ 
by $E_{pp'}$ or, equivalently, $E_{qq'}$. 

We stress that the quasi-particle approximation implies that the energy
splitting $E_{nn'}$ is large compared with $\tilde K$. We have also assumed
that $\tilde K$ is small and only expanded $\tilde K$ to leading order, i.e.,
the Born approximation. To higher order in the system-bath coupling
the diagrammatic contributions to $\tilde K$ give rise to further corrections
to the Green's functions $G_{R,A}$ describing the propagation of the system
and vertex corrections to the interaction $H_{SB}$. To fourth order
we thus obtain twelve diagrams involving two bath correlation functions,
and the application of the quasi-particle approximation becomes
more complex. The higher order corrections deserve further studies but
will not be considered here.

\subsection{Redfield and Lindblad equations}
Implementing the Markov approximation by choosing $\omega=E_{qq'}$ in
(\ref{kernelborn}) we obtain the standard Redfield kernel, see e.g.  
\cite{Purkayastha23},
\begin{eqnarray}
\tilde K_{pp',qq'}=
&&-\frac{1}{2}\delta_{p'q'}\sum_{\alpha\beta l}S_{pl}^\alpha S_{lq}^\beta
\tilde D^{\alpha\beta}(E_{ql})~~~~~~~(1)
\nonumber
\\
&&-\frac{1}{2}\delta_{pq}\sum_{\alpha\beta l }S_{q'l}^\alpha S_{lp'}^\beta
\tilde D^{\alpha\beta}(E_{q'l})~~~~~~(2)
\nonumber
\\
&&+\frac{1}{2}\sum_{\alpha\beta }S_{pq}^\beta S_{q'p'}^\alpha 
\tilde D^{\alpha\beta}(E_{q'p'})~~~~~~~~(3)
\nonumber
\\
&&+\frac{1}{2}\sum_{\alpha\beta }S_{pq}^\beta S_{q'p'}^\alpha
\tilde D^{\alpha\beta}(E_{qp});~~~~~~~~(4)
\label{redkernel1}
\end{eqnarray}
the numbering above refers to the numbered diagrams in Fig. 1e.

In order to obtain the Lindblad kernel and thus obtain consistency with
a quantum map, we introduce the system operators, see also \cite{Breuer06}, 
\begin{eqnarray}
&&S^\alpha_{pq}(\omega)=S^\alpha_{pq}\delta(\omega-E_{qp}),
\label{s1}
\\
&&S^{\alpha\dagger}_{pq}(\omega)=
S^{\alpha\dagger}_{pq}\delta(\omega-E_{pq}),
\label{s11}
\\
&&S^\alpha_{pq}=\int^{\infty}_{-\infty} d\omega S^\alpha_{pq}(\omega),
\label{s2}
\\
&&S^{\alpha\dagger}_{pq}=
\int^{\infty}_{-\infty} d\omega S^{\alpha\dagger}_{pq}(\omega),
\label{s22}
\end{eqnarray}
corresponding to a projection onto the energy eigenspace,
and the bath correlations
\begin{eqnarray}
\tilde\gamma^{\alpha\beta}(\omega)=
\langle B^{\alpha\dagger}B^\beta\rangle(\omega).
\end{eqnarray}
By insertion in (\ref{redkernel1}), using $H_{SB}=H_{SB}^\dagger$, i.e.,
$\sum_\alpha S^\alpha B^\alpha=
\sum_\alpha S^{\alpha\dagger} B^{\alpha\dagger}$, we thus obtain
\begin{eqnarray}
\tilde K_{pp',qq'}=
&&-\frac{1}{2}
\delta_{p'q'}\int^{\infty}_{-\infty} d\omega d\omega'\sum_{\alpha\beta l}
S_{pl}^{\alpha\dagger}(\omega')
S_{lq}^\beta(\omega)
\tilde\gamma^{\alpha\beta}(\omega)
\nonumber
\\
&&-\frac{1}{2}
\delta_{pq}\int^{\infty}_{-\infty} d\omega d\omega'\sum_{\alpha\beta l }
S_{q'l}^{\alpha\dagger}(\omega)
S_{lp'}^\beta(\omega')
\tilde\gamma^{\alpha\beta}(\omega)
\nonumber
\\
&&+\frac{1}{2}
\int^{\infty}_{-\infty} d\omega d\omega'\sum_{\alpha\beta }
S_{pq}^\beta(\omega') 
S_{q'p'}^{\alpha\dagger}(\omega) 
\tilde\gamma^{\alpha\beta}(\omega)
\nonumber
\\
&&+\frac{1}{2}
\int^{\infty}_{-\infty} d\omega d\omega'\sum_{\alpha\beta }
S_{pq}^\beta(\omega)
S_{q'p'}^{\alpha\dagger}(\omega')
\tilde \gamma^{\alpha\beta}(\omega).
\label{redkernel2}
\end{eqnarray}
The final step is to impose the rotating wave approximation (RWA),
i.e. discarding rotating terms in the long time limit. Here it corresponds
to imposing the delta function constraint $\delta(\omega-\omega')$,
corresponding 
to $\lim_{t\to\infty}\exp(t(\omega-\omega'))\sim\delta(\omega-\omega')$.
By insertion in (\ref{redkernel2}) we obtain the Lindblad kernel
\begin{eqnarray}
\tilde K_{pp',qq'}=
&&-\frac{1}{2}
\delta_{p'q'}\int^{\infty}_{-\infty}d\omega\sum_{\alpha\beta l}
S_{pl}^{\alpha\dagger}(\omega)
S_{lq}^\beta(\omega)
\tilde\gamma^{\alpha\beta}(\omega)
\nonumber
\\
&&-\frac{1}{2}
\delta_{pq}\int^{\infty}_{-\infty} d\omega\sum_{\alpha\beta l }
S_{q'l}^{\alpha\dagger}(\omega)
S_{lp'}^\beta(\omega)
\tilde\gamma^{\alpha\beta}(\omega)
\nonumber
\\
&&+\int^{\infty}_{-\infty} d\omega\sum_{\alpha\beta }
S_{pq}^\beta(\omega) 
S_{q'p'}^{\alpha\dagger}(\omega) 
\tilde\gamma^{\alpha\beta}(\omega),
\label{lindblad}
\end{eqnarray}
or in operator form the Lindblad master equation in the first standard
form
\begin{eqnarray}
\frac{d\rho_S(t)}{dt}=&&-i[H_S,\rho_S(t)] 
\nonumber
\\
&&+\sum_{\alpha\beta}\int^{\infty}_{-\infty} d\omega\tilde\gamma^{\alpha\beta}(\omega)
\bigg [S^\beta(\omega)\rho_S(t)S^{\alpha\dagger}(\omega)-\frac{1}{2}
\{S^{\alpha\dagger}(\omega)S^\beta(\omega),\rho_S(t) \}\bigg ].
\label{linmaster}
\end{eqnarray}
Referring to the discussion in \cite{Breuer06} the bath correlation
function $\tilde\gamma^{\alpha\beta}(\omega)=
\langle B^{\alpha\dagger}B^\beta\rangle(\omega)$ forms a
positive matrix and diagonalisation yields the standard Lindblad
equation.
\section{\label{new}Present work}
In this section we present our new results referring to the summary in 
Sec. \ref{review}.
\subsection{Discrepancy in the standard Redfield equation}
The standard derivation of the Redfield equation \cite{Breuer06,Purkayastha23} 
yields in operator form the Markov equation
\begin{eqnarray}
&&\frac{d\rho_S(t)}{dt}=-i[H_S,\rho_S(t)] +K\rho_S(t),
\label{redmaster}
\end{eqnarray}
where the Redfield kernel is given by
\begin{eqnarray}
K\rho_S(t)=
&&-\sum_{\alpha\beta}
\int^{\infty}_{-\infty} dt'S^\alpha G_R(t',0)S^\beta G_A(0,t')\rho_S(t)
D^{\alpha\beta}(t',0)
\nonumber
\\
&&-\sum_{\alpha\beta}
\int^{\infty}_{-\infty} dt'\rho_S(t)G_R(t',0)S^\alpha G_A(0,t')S^\beta
D^{\alpha\beta}(0,t')
\nonumber
\\
&&+\sum_{\alpha\beta}
\int^{\infty}_{-\infty} dt'S^\beta\rho_S(t) G_R(t',0)S^\alpha G_A(0,t')
D^{\alpha\beta}(0,t')
\nonumber
\\
&&+\sum_{\alpha\beta}
\int^{\infty}_{-\infty} dt'G_R(t',0)S^\beta G_A(0,t')\rho_S(t)S^\alpha
D^{\alpha\beta}(t',0).
\label{redkernel3}
\end{eqnarray}
Inserting $G_{R,A}$ from (\ref{greenret} - \ref{greenadv}) and the bath 
correlation function, $\tilde D^{\alpha\beta}(\omega)$ from 
(\ref{bath} - \ref{fourier-bath}), keeping only the dissipative part, we obtain 
in the energy basis the Redfield kernel in (\ref{redkernel1}).
By inspection we note that the trace condition 
$\text{Tr}(\tilde K) = \sum_p\tilde K_{pp,qq'}=0$
is satisfied, yielding a constant $\sum_n\rho_S(t)_{nn}$, i.e. conservation of
probability.

Next we turn to a discussion of the individual terms in the Redfield
kernel in (\ref{redkernel1}). Referring to the diagrams (1) and (2), the 
first two terms in (\ref{redkernel1}) describe the combined
transitions in $S_{pl}^\alpha S_{lq}^\beta$ and 
$S_{q'l}^\alpha S_{lp'}^\beta$ associated with an exchange of the 
energy quanta (photons or phonons) $E_{ql}$ and $E_{q'l}$ with the bath, 
respectively. These processes correspond to a correction of the 
system propagation given by the Green's functions $G_R$ and $G_A$. 

In a similar manner the diagrams (3) and (4) and the two last terms 
in (\ref{redkernel1}) corresponding 
to the transitions $S_{pq}^\beta S_{q'p'}^\alpha$ are associated with 
the energy transitions $E_{q'p'}$ and $E_{qp}$, respectively. These 
processes correspond to cross correlations between the forward 
and backward propagation in the evolution of the density operator 
$\rho_S(t)$.

Thus we note a clear discrepancy in the derivation of the Redfield equation. 
In the terms (1) and (2) the exchanged quanta are only associated
with a single matrix element. In (1) the quantum $E_{ql}$ is associated
with $S_{lq}^\beta$ in (2) $E_{q'l}$ with $S_{q'l}^\alpha$. In a similar
manner in (3) $E_{q'p'}$ is related to  $S_{q'p'}^\alpha$ and in (4) 
$E_{qp}$ with $S_{pq}^\beta$. 

Referring to the field theoretical analysis and quasi-particle 
approximation in Sec. \ref{review}, the ansatz $\omega=E_{qq'}$ in 
(\ref{kernelborn}) precisely yields the Redfield kernel in 
(\ref{redkernel1}). On the other hand, the equivalent ansatz 
$\omega=E_{pp'}$ yields a Redfield equation where by 
 inspection the transmitted quanta are associated with the 
 other matrix element in the bilinear combination.
 \subsection{Resolving the discrepancy - Energy conservation}
The issue is resolved by noting that the quasi-particle approximation 
implies the equivalence of the assignments $\omega=E_{pp'}$ and 
$\omega=E_{qq'}$ and thus imposes a selection rule on the contributions to 
the Redfield kernel. In order to resolve the inconsistency
issue, we are thus led to impose energy conservation in the individual 
contributions to the Born approximation. In a quantum field context this 
corresponds to "on the mass shell'' terms, 
an issue which will be considered in more detail
in the discussion section.
 In term (1) we have 
$E_{pl}=E_{ql}$, i.e., $E_p=E_q$, in term (2) $E_{p'l}=E_{q'l}$,
i.e., $E_{p'}=E_{q'}$; in terms (3) and (4) we infer $E_{pq}=E_{p'q'}$. 
Thus incorporating energy conservation on the Born level by means of 
a selection rule in terms of delta function constraints we obtain the 
modified Redfield kernel
\begin{eqnarray}
\tilde K_{pp',qq'}=
&&-\frac{1}{2}\delta_{p'q'}\sum_{\alpha\beta l}S_{pl}^\alpha S_{lq}^\beta
\tilde D^{\alpha\beta}(E_{ql})\delta(E_{pl}+E_{lq})
\nonumber
\\
&&-\frac{1}{2}\delta_{pq}\sum_{\alpha\beta l }S_{q'l}^\alpha S_{lp'}^\beta
\tilde D^{\alpha\beta}(E_{q'l})\delta(E_{q'l}+E_{lp'})
\nonumber
\\
&&+\sum_{\alpha\beta }S_{pq}^\beta S_{q'p'}^\alpha 
 \tilde D^{\alpha\beta}(E_{q'p'})\delta(E_{pq}+E_{q'p'});
\label{redkernel4}
\end{eqnarray}
we have here combined the two last terms.
 \subsection{Deriving the Lindblad equation without the RWA}
The standard derivation of the Lindblad equation \cite{Breuer06} yields in 
operator form the Markov master equation in (\ref{linmaster}), corresponding
to the Lindblad kernel in the first standard form
\begin{eqnarray}
\tilde K\rho_S(t)=&&\sum_{\alpha\beta}\int^{\infty}_{-\infty} d\omega
\tilde\gamma^{\alpha\beta}(\omega)\bigg[S^\beta(\omega)\rho_S(t) S^{\alpha\dagger}(\omega)-
\frac{1}{2}\{S^{\alpha\dagger}(\omega)S^\beta(\omega),\rho_S(t) \}\bigg],
\label{linkernel1}
\end{eqnarray}
where $\tilde\gamma^{\alpha\beta}(\omega)=
\langle B^{\alpha\dagger}B^\beta\rangle(\omega)$. 
Expanding in an energy basis we have
\begin{eqnarray}
&&\tilde K_{pp',qq'}=
\nonumber
\\
&&\sum_{\alpha\beta}
\int^{\infty}_{-\infty} d\omega\tilde\gamma^{\alpha\beta}(\omega)\bigg[S_{pq}^\beta(\omega)S_{q'p'}^{\alpha\dagger}(\omega)-
\frac{1}{2}\sum_l
\Big(\delta_{p'q'}S_{pl}^{\alpha\dagger}(\omega)S_{lq}^\beta(\omega)
+\delta_{pq}S_{q'l}^{\alpha\dagger}(\omega)S_{lp'}^\beta(\omega)\Big)\bigg].
\nonumber
\\
\label{linkernel2}
\end{eqnarray}
Introducing $S^\alpha_{pq}(\omega)=S^\alpha_{pq}\delta(\omega-E_{qp})$,
$S^{\alpha\dagger}_{pq}(\omega)=
S^{\alpha\dagger}_{pq}\delta(\omega-E_{pq})$, integrating over $\omega$ 
and using $\sum_\alpha S^\alpha B^\alpha= 
\sum_\alpha S^{\alpha\dagger} B^{\alpha\dagger}$ and 
$\tilde D^{\alpha\beta}(\omega)=\langle B^{\alpha}B^\beta\rangle(\omega)$,
we obtain complete agreement with the modified Redfield equation in
(\ref{redkernel4}). 

It is an important result of the present analysis that imposing a selection
rule on the individual terms in the perturbation expansion and thus enforcing
energy conservation automatically yields the Lindblad equation. Since the 
Lindblad equation also follows from applying the the rotating wave approximation
to the Redfield equation, discarding oscillating terms, the two methods are
equivalent. In our view the selection rule argument represents a more fundamental 
insight in line with a field theoretical approach. 

Finally, we note that the RWA approximation in the dipole approximation
in the case of an atom interacting with the radiation field can result in
nonlocal interaction \cite{Clerk98}, see also \cite{Jorgensen22} for the
breakdown of the RWA in superradiance
\section{\label{discussion} Summary and Discussion}
\subsection{Summary}
The purpose of the present work together with the work in 
\cite{Fogedby22} is to apply field theoretical condensed
matter methods to open quantum systems. 1) Assuming a multi-oscillator
heat bath we derive a Dyson equation for the transmission matrix
describing the evolution of the system density matrix. The Dyson equation
includes secular effects and replaces the physical assumption
$\rho(t)=\rho_S(t)\rho_B$ used in the standard derivation in order
to obtain closure. 2) The Dyson equation in turn yields a general
non markovian master equation for the system density operator,
independent of the initial preparation. 3) The kernel in the master
equation allows for an expansion in powers of the system-bath
interaction, characterised by Feynman-type diagrams. 4) In the
Born approximation the kernel is given by four diagrams involving
one bath correlation function (fourth order is given by twelve
diagrams including two correlation functions). 5) Applying the
quasi-particle approximation, corresponding to a time-scale
separation, we obtain the markovian Redfield master equation.
6) Imposing energy conservation as a selection rule on the 
individual contributions following from an ambiguity in the
quasi-particle assignment  we obtain the Lindblad equation.

We stress again that the field theoretical approach 
is here limited to the weak coupling case, a multi-oscillator
heat bath,  and large energy splittings compared to
the system-bath coupling.
\subsection{Quantum field theory versus open quantum systems}
There is an interesting parallel 
between quantum field theory (QFT)
and open quantum systems (OQS).  In QFT the elementary particles
interact with a vacuum, in OQS the quantum system interacts with
a heat bath. In QFT there are vacuum fluctuations, in OQS there
are temperature fluctuations. In QFT perturbation theory a
particle is described by the Green's function $G(E,p)=
1/(E-p^2/2m)$, in OQS perturbation theory the quantum system
is described by the Green's function $G(\omega,E)=1/(\omega-E)$.
Vacuum fluctuations take place on a fast time scale $\Delta t$, 
likewise, temperature fluctuations also take place on a fast time scale $\Delta t$. 
The uncertainty relation $\Delta E\Delta t\sim\hbar$ implies
that in QFT the virtual particle in perturbation theory can exist off the mass shell
$E=p^2/2m$ for a short time. Likewise, for a heat bath we
suggest that the quantum system can exist off the mass shell
$\omega=E$. In scattering experiments in QFT (the S-matrix)
the real particles are free and must live on the mass shell. 
Likewise, in OQS at long times in the Markov regime the
system must be confined to the mass shell, i.e., 
$\omega$ locked on to $E$ in order to properly represent
the system density matrix. This is the qualitative  basis for the
quasi-particle approximation.
\subsection{Populations and Coherences in the steady state}
Both the Redfield and Lindblad master equations satisfy the trace 
condition and thus the conservation of probability but only the 
Lindblad equation corresponds to a quantum map and thus
guarantees positive probabilities. Given the explicit
expression in an energy basis of the modified Redfield kernel
in (\ref{redkernel4}), shown above to be equivalent to the Lindblad
kernel (\ref{linkernel2}), we discuss below populations and 
coherences in the steady state. 
\subsubsection{\bf\em General properties}
From the master equation in (\ref{master}) in the Markov limit,
$K(t,t')_{pp',qq'}\to K_{pp',qq'}\delta(t-t')$, the steady state is obtained
by setting $d\rho_S(t)_{pp'}/dt=0$. Denoting the steady state density
matrix by $(\rho_S^{st})_{pp'}$ 
and noting that  $E_{pp'}=E_p-E_{p'}$, we obtain the general condition
\begin{eqnarray}
-iE_{pp'}(\rho_S^{st})_{pp'}+
\sum_{qq'}\tilde K_{pp',qq'}(\rho_S^{st})_{qq'}=0. 
\label{steady}
\end{eqnarray}
Populations are determined by the diagonal elements 
$(\rho_S^{st})_{pp}$,  
\begin{eqnarray}
\sum_{q}\tilde K_{pp,qq}(\rho_S^{st})_{qq}+
\sum_{q\neq q'} \tilde K_{pp,qq'}(\rho_S^{st})_{qq'}=0. 
\label{pop}
\end{eqnarray}
Coherences are given by the off-diagonal elements 
$(\rho_S^{st})_{pp'}$ for $p\neq p'$, 
\begin{eqnarray}
-iE_{pp'}(\rho_S^{st})_{pp'}+
\sum_{q} \tilde K_{pp',qq}(\rho_S^{st})_{qq}+
\sum_{q\neq q'} \tilde K_{pp',qq'}(\rho_S^{st})_{qq'} = 0. 
\label{coh}
\end{eqnarray}
By inspection of (\ref{pop}) and (\ref{coh}) we infer the following:

a) In the case where $\tilde K_{pp,qq'}=0$ for 
$q\neq q'$ the populations form a block and do not couple to the 
coherences; for $\tilde K_{pp,qq'}\neq 0$ for $q\neq q'$ coherences and 
populations interact and coherences can contribute to the populations.

b) In the case where $\tilde K_{pp',qq}=0$ for $p\neq p'$ the coherences
form a block not interacting with the populations; for $\tilde K_{pp',qq}\neq0$
coherences couple to the populations.
\subsubsection{\bf\em Modified Redfield kernel (Lindblad kernel)}
\paragraph{Non-degenerate case:}
From the modified Redfield kernel in (\ref{redkernel4}) we have
$\delta(E_{pl}+E_{lq})=\delta(E_{pq})$ and 
$\delta(E_{q'l}+E_{lp'})=\delta(E_{q'p'})$ implying in the non-degenerate
case $\delta_{pq}$ and $\delta_{p'q'}$, respectively,  and we obtain
\begin{eqnarray}
\tilde K_{pp',qq'}=
&&-\frac{1}{2}\delta_{pq}\delta_{p'q'}\sum_{\alpha\beta l}
(S_{pl}^\alpha S_{lq}^\beta\tilde D^{\alpha\beta}(E_{ql})
+S_{q'l}^\alpha S_{lp'}^\beta
\tilde D^{\alpha\beta}(E_{q'l}))
\nonumber
\\
&&+\sum_{\alpha\beta }S_{pq}^\beta S_{q'p'}^\alpha 
 \tilde D^{\alpha\beta}(E_{q'p'})\delta( E_{pq}+E_{q'p'}).
\label{redkernelnd}
\end{eqnarray}
In the population sector for $p=p'$ inspection of (\ref{redkernelnd})
implies $q=q'$ and from $\delta(E_{pq}+E_{q'p})=\delta(E_{q'q})=
\delta_{q'q}$ and
we obtain the non vanishing term
\begin{eqnarray}
\tilde K_{pp,qq}=
&&-\delta_{pq}\sum_{\alpha\beta l}
S_{pl}^\alpha S_{lp}^\beta\tilde D^{\alpha\beta}(E_{pl})
+\sum_{\alpha\beta }S_{pq}^\beta S_{qp}^\alpha 
 \tilde D^{\alpha\beta}(E_{qp}).
\label{redkernelpop}
\end{eqnarray}
We conclude that the population contributions are confined to 
the population block.

In the coherence sector for $p\neq p'$ the delta function constraints
in the first term in (\ref{redkernelnd}) implies $q\neq q'$ with a
remaining $\delta_{pq}\delta_{p'q'}$. For $p\neq p'$ the delta function 
constraint in the last term $\delta(E_{pq}+E_{q'p'})
=\delta(E_{pp'}-E_{qq'})$ also implies $q\neq q'$. As a consequence 
the coherence contributions are confined to the coherence block.

We conclude that in the non-degenerate case the population block
and coherence block are uncoupled. The contributions to the kernel
$\tilde K_{pp',qq'}$ will depend on the model under investigation. 
In an actual calculations only the populations persist in the steady 
state, the coherences form transients and die out.
\paragraph{Degenerate case:}
In the degenerate case $\delta(E_{pq})$ or $\delta(E_{p'q'})$
do not imply $p=q$ or $p'=q'$ since the states $p$ and $q$
or $p'$ and $q'$ might be degenerate and have the same energies.

In the population sector for $p=p'$ we have from 
the kernel in (\ref{redkernel4})
\begin{eqnarray}
\tilde K_{pp,qq'}=
&&-\frac{1}{2}\delta_{pq'}\sum_{\alpha\beta l}S_{pl}^\alpha S_{lq}^\beta
\tilde D^{\alpha\beta}(E_{ql})\delta(E_{pq})
\nonumber
\\
&&-\frac{1}{2}\delta_{pq}\sum_{\alpha\beta l }S_{q'l}^\alpha S_{lp}^\beta
\tilde D^{\alpha\beta}(E_{q'l})\delta(E_{pq'})
\nonumber
\\
&&+\sum_{\alpha\beta }S_{pq}^\beta S_{q'p}^\alpha 
 \tilde D^{\alpha\beta}(E_{q'p})\delta(E_{q'q}),
\label{redkernel-6}
\end{eqnarray}
which shows that the populations couple to coherences in the
case where the states $q$ and $q'$ are degenerate yielding
$E_{qq'}=0$. 
\subsubsection{\bf\em Qubit case}
In the case of a non degenerate single qubit or two level system
with $H_S=(\omega_0/2)\sigma^z$ and interaction 
$H_{SB}=\sigma^+ B+\sigma^-B^\dagger$, where 
$\sigma^z,\sigma^+,\sigma^-$ are the Pauli matrices, $\omega_0$ the 
level splitting and $B$ the bath operator, we chose in accordance
with $H_{SB}=\sum_\alpha S^\alpha B^\alpha$ the assignment
$S_1=\sigma^+, S_2=\sigma^-, B_1=B,B_2=B^\dagger$.

In this case the non vanishing matrix elements are the diagonal elements 
$K_{pp,qq}$ and the off-diagonal elements $K_{+-,+-}$ and $K_{-+,-+}$. 
The diagonal and off-diagonal elements are uncoupled. 
$K_{pp,qq}$ yield the populations in the stationary state. 
$K_{+-,+-}$ and $K_{-+,-+}$ give rise to transient coherences 
vanishing in the stationary state. By inspection the selection rules are 
automatically satisfied  and the Redfield and Lindblad equations 
are equivalent.
\subsection{Field theoretical approach compared with EQM, NZ and TCL}
In open quantum systems the standard derivations of the Redfield
and Lindblad master equations are based on three approaches:
The equation of motion approach (EQM), the Nakajima-Zwanzig 
projection operator approach (NZ) and the time-convolutionsless 
approach (TCL).
Below we attempt to demonstrate how the field theoretical approach 
(FTM) compares with
the EQM, the NZ and the TCL.  In Appendix \ref{appendix}
we have summarised aspects of the EQM, the NZ and the TCL methods.
\subsubsection{FTM versus EQM}
The EQM is based on the von Neumann equation for the total
density operator $\rho(t)$. Solving the equation of motion to 
second order, reinserting and tracing over the bath one 
obtains an equation of motion for the density operator $\rho_S(t)$,
depending on $\rho(t')$ and the initial condition at $t=t_i$. 
Subsequently, assuming a weak coupling to the bath one makes the
approximation or ansatz $\rho(t')\approx\rho_S(t')\rho_B$ in order to obtain 
closure. Assuming a time scale separation the Markov limit 
is obtained by replacing  $\rho_S(t')$ with $\rho_S(t)$. Moreover,
the dependence on of the initial condition is removed
by a change of variables and letting $t_i\to -\infty$.

In the FTM the Dyson equation for the transmission operator $T$,
based on a multi-oscillator bath, includes secular effects
and readily yields a proper master equation of motion 
not depending on the initial condition. Moreover, the Markov limit 
is obtained by applying the quasi-particle approximation.

At this stage both the EQM and the FTM yield the Redfield
equation. In order to obtain the Lindblad equation the RWA
is applied in the EQM; in the FTM energy conservation on
the Born level, corresponding to selection rules, yields the Lindblad equation
\subsubsection{FTM versus NZ}
The NZ is based on a projection method. Introducing the
projection operator $P$ according to the definition 
$P\rho(t)=\rho_S(t)\rho_B$ one basically formalises the EQM.
The operator $P$ projects onto the quantum system characterised by 
$\rho_S(t)$, whereas the operator $Q=I-P$
projects onto the bath. To second order the NZ is
equivalent to the EQM and yields the Redfield equation.
The FTM is based on the multi-oscillator bath, the Dyson equation
and the quasi-particle approximation.
\subsubsection{FTM versus TCL}
An extension of the NZ yields the TCL allowing for a perturbations 
expansion in powers of the coupling between the system and bath.
The TCL expansion operates in the Markov regime by introducing
forward and backward propagation in terms of appropriate
Green's functions and in this manner directly
yields Markov master equations. To second order one obtains
the Redfield equation. 
Assuming a coupling of the dipole form 
$H_{SB}=\sum_\alpha S^\alpha B^\alpha$ the TCL expansion
is based on a Gaussian bath and a cumulant expansion.

In the FTM the perturbation expansion operates in the non-Markovian
regime and applies to the kernel in the master equation.
The derivation of the Dyson equation is based on the Keldysh 
contour from the initial time $t_i$ to the time $t$ and back.
This particular feature bears a resemblance to the procedure
in the TCL. Moreover, the assumption of a Gaussian bath in the
TCL is in the FTM exemplified by the multi-oscillator bath
together with Wick's theorem which appears to correspond to
the cumulant expansion in the TCL.
A more detailed comparison of the TCL and the FTM lies
beyond the scope of the present work.
\subsection{Recent work}
\subsubsection{Universal Lindblad Equation (ULE)}
In order to circumvent the well-known issues regarding the Redfield
equation (negative probabilities) and the Lindblad equation 
(weak coupling, large energy splitting) a generalisation of the
Lindblad equation,  {\em The Universal Lindblad Equation} (ULE),
has been proposed \cite{Nathan20,Nathan21,Nathan18}. The ULE has
the Lindblad form, guarantees a positive and trace-preserving
density operator and allows for a study of a 
wide range of systems in the context of quantum entanglement, quantum
thermodynamics and quantum computers . While the ULE can be applied to
a variety of systems not accessible by the Redfield or Lindblad equations,
it seems to suffer from some fundamental limitations: 1) It
does not relax to the true Gibbs state \cite{Lee20,Nathan22}, 2) it generates
spurious currents in the steady state state \cite{Tupkary22}.
For further references see \cite{Kirsanskas18,Nathan19,Davidovich20}.


\subsubsection{Keldysh field theory (KFT) }
The KFT is based on functional path integrals with a closed time path
and a Keldysh action combined with a diagrammatic expansion.
It differs from the present approach which is based on a direct expansion
of the evolution of the density operator and a general master equation.

Here we list recent work where field theoretical methods are used.
In \cite{McDonald23} the Keldysh field theoretical techniques 
developed for non equilibrium many body systems is applied to 
open quantum systems. In \cite{Reimer19a} the Keldysh method
is used to investigate positivity of the quantum map in open
systems. In \cite{Thompson23} Keldysh field theory is applied
to Lindblad dynamics in open systems. In \cite{Sieberer16}
Keldysh theory is applied to driven open systems.
Aspects of the work in \cite{Gu24} based on diagrammatics
bears some resemblance to the present work, however,
in \cite{Gu24} the focus is on the TCL. Likewise, in 
\cite{Ferguson21} the analysis makes use of the Keldysh
contour as in the present work but the focus is the TCL.
In \cite{Nestmann21a} time-local and time-nonlocal perturbation
expansions are analysed making use of diagrammatics.
In \cite{Nestmann21b} the NZ and TCL and their interconnection
is studied.
\section{\label{appendix}Appendix}
\subsection{Deriving the transmission matrix}
The density operator for the total system  evolves in time according to 
\begin{eqnarray}
&&\rho(t)=U(t,t_i)\rho(t_i)U(t,t_i)^\dagger,
\label{a1}
\\
&&U(t,t')=\exp(-iH(t-t')),
\label{a2}
\end{eqnarray}
where $t_i$ is the initial time. Since $\text{Tr}\rho(t)=\text{Tr}\rho(t_i)$ 
the trace condition $\text{Tr}\rho(t)=1$ is preserved at all times.

The open quantum system is described by  $\rho_S(t)$ obtained by 
tracing over the bath degrees of freedom, i.e., 
$\rho_S(t)=\text{Tr}_B\rho(t)$. Since $\text{Tr}=\text{Tr}_S\text{Tr}_B$ we
infer the trace condition $\text{Tr}_S\rho_S(t)=1$.  A fundamental 
issue 
is  the time evolution of the reduced system density operator 
$\rho_S(t)$.
Regarding a systematic approach the natural expansion parameter 
is given
by the system-bath coupling $H_{SB}$. Introducing the Hamiltonian 
for the
combined system and bath $H_0=H_S+H_B$ and applying the 
interaction 
representation we obtain the evolution operator
\begin{eqnarray}
&&U(t,t')=\exp(-iH_0t)\bigg(\exp\bigg[-i\int_{t'}^tdt''H_{SB}(t'')\bigg]\bigg)_+\exp(iH_0t'),
\label{a3}
\\
&&H_{SB}(t)=\exp(iH_0t)H_{SB}\exp(-iH_0t),
\label{a4}
\end{eqnarray}
where the time ordered products $(\cdots)_+$ is shorthand for
the expansion
\begin{eqnarray}
\Big(\cdots\Big)_+=
\sum_{n=0}(-i)^n\int_{t'}^{t}dt_{n}\int_{t'}^{t_n}dt_{n-1}\cdots\int_{t'}^{t_2}dt_{1}
H_{SB}(t_{n})\cdots H_{SB}(t_{1}).
\label{a5}
\end{eqnarray}
Introducing the retarded and advanced Green's functions
\begin{eqnarray}
&&G_R(t,t')=-i\eta(t-t')\exp(-iH_S(t-t')), \label{a6}
\\
&&G_A(t,t')=+i\eta(t'-t)\exp(-iH_S(t-t')),
\label{a7}
\end{eqnarray}
inserting $H_{SB}=\sum_\alpha S^\alpha B^\alpha$ and 
using the notation $S^\alpha_n(t_n)=
\exp(iH_St_n)S^\alpha_n\exp(-iH_St_n)$ and $B^\alpha_n(t_n)=
\exp(iH_Bt_n)B^\alpha_n\exp(-iH_Bt_n)$ we obtain
\begin{eqnarray}
&&U(t,t')=
\nonumber
\\
&&i\sum_{n=0}\int^{\infty}_{-\infty}dt_{n}dt_{n-1}\cdots dt_{1}G_R(t,t_{n})S_{n}
G_R(t_n,t_{n-1})S_{n-1}\cdots S_2G_R(t_2,t_{1})S_1G_R(t_1,t')\times
\nonumber
\\
&&\exp(-iH_Bt)B_{n}(t_n)B_{n-1}(t_{n-1})\cdots B_2(t_2)B_1(t_1)\exp(iH_Bt'),
\label{a8}
\end{eqnarray}
and a corresponding expression for $U(t,t')^\dagger$;
for clarity we have used the notation $H_{SB}=SB$.

The time evolution of the density operator $\rho_S(t)$ is then given by the
double sum  (the so-called Keldysh contour from time $t$ to time $t_i$ and
back to time $t$),
\begin{eqnarray}
\rho_S(t)=
\sum_{n=0,m=0}
&&\int^{\infty}_{-\infty} dt_n\cdot\cdot dt_1G_R(t,t_n)S_nG_R(t_n,t_{n-1})S_{n-1}
\cdot\cdot S_2G_R(t_2,t_{1})S_1G_R(t_1,t_i)
\rho_S(t_i)\times
\nonumber
\\
&&\int^{\infty}_{-\infty} du_1\cdot\cdot du_m
G_A(t_i,u_1)S_1 G_A(u_1,u_2)S_2 \cdot\cdot 
S_{m-1} G_A(u_{m-1},u_m)S_mG_A(u_m,t)\times
\nonumber
\\
&&\text{Tr}_B[\rho_B B_1(u_1)\cdot\cdot
B_m(u_m)B_n(t_n)\cdot\cdot B_1(t_1)],
\label{a9}
\end{eqnarray}
where we have traced over the bath variables $B^\alpha_n(t_n)$ with
weight given by the bath density operator $\rho_B$. 

The expression (\ref{a9}) involves the bath correlation function 
$\text{Tr}_B[\rho_B B_1\cdots B_n]\equiv\langle B_1\cdots B_n\rangle$. Choosing for the bath a collection
of independent quantum oscillators according to the Caldeira-Leggett 
prescription \cite{Caldeira83a,Caldeira83b} and invoking 
Wick theorem \cite{Gaudin60} the bath trace breaks up in the product of all 
possible pairings $\langle B_pB_q\rangle$, e.g.
 $\langle B_1B_2B_3B_4\rangle=
\langle B_1B_2\rangle\langle B_3B_4\rangle+
\langle B_1B_3\rangle\langle B_2B_4\rangle+
\langle B_1B_4\rangle\langle B_2B_3\rangle$; we here assume that 
$\langle B_n\rangle=0$. 

Introducing the transition operator $T(t,t')$ and Green function identities 
according to
\begin{eqnarray}
&&\rho_S(t)=T(t,t_i)\rho_S(t_i),
\label{a10}
\\
&&G_R(t,t')=iG_R(t,t'')G_R(t'',t'),
\label{a11}
\\
&&G_A(t,t')=-iG_A(t,t'')G_A(t'',t'),
\label{a12}
\end{eqnarray}
inspection of (\ref{a9}) yields the Dyson equation
\begin{eqnarray}
T(t,t')=T^0(t,t')+\int^{\infty}_{-\infty} dt'' dt''' T^0(t,t'')K(t'',t''')T(t''',t'),
\label{a13}
\end{eqnarray}
where the unperturbed operator $T^0$ is given by
\begin{eqnarray}
T^0(t,t')=G_R(t,t')G_A(t',t).
\label{a14}
\end{eqnarray}
The  kernel $K(t,t')$ is given as an expansion in powers of
$H_{SB}$ involving the free propagators $G_R(t,t')$ 
and $G_A(t,t')$,
the system operators $S^\alpha_n$, and the bath correlations 
$\langle B^\alpha_p(t_p)B^\beta_q(t_q)\rangle$.

The Dyson equation in (\ref{a13}) implies  an integral
equation for $\rho_S$, 
\begin{eqnarray}
\rho_S(t)= T^0(t,t_i)\rho_S(t_i)+\int^{\infty}_{-\infty} dt' dt'' T^0(t,t')K(t',t'')\rho_S(t'').
\label{a15}
\end{eqnarray}
Finally , using $dT^0(t,t')/dt=-i[H_S,T^0(t,t')] + \delta(t-t')$ we 
infer a master equation for $\rho_S(t)$
\begin{eqnarray}
\frac{d\rho_S(t)}{dt}=-i[H_S,\rho_S(t)]+\int^{\infty}_{-\infty} dt' K(t,t')\rho_S(t').
\label{a16}
\end{eqnarray}
To second order in the interaction $H_{SB}$ corresponding to the
Born approximation the kernel $K(t,t')$ is given by
\begin{eqnarray}
K(t,t')\rho_S(t)=
&&-\sum_{\alpha\beta}S^\alpha G_R(t,t')S^\beta\rho_S(t) G_A(t',t)
D^{\alpha\beta}(t,t')
\nonumber
\\
&&-\sum_{\alpha\beta}G_R(t,t')\rho_S(t) S^\alpha G_A(t',t)S^\beta
D^{\alpha\beta}(t',t)
\nonumber
\\
&&+\sum_{\alpha\beta}S^\beta G_R(t,t')\rho_S(t) S^\alpha G_A(t',t)
D^{\alpha\beta}(t',t)
\nonumber
\\
&&+\sum_{\alpha\beta}G_R(t,t')S^\beta\rho_S(t) G_A(t',t)S^\alpha
D^{\alpha\beta}(t,t').
\label{a17}
\end{eqnarray}
In the energy basis and in Fourier space (\ref{a17}) yields 
(\ref{kernelborn}).
\subsection{Equation of motion approach (EQM)} 
In the EQM the starting point is the von Neumann equation of motion 
in the interaction representation for the global density operator 
$\rho(t)$,
\begin{eqnarray}
\frac{d\rho(t)}{dt}=-i[H_{SB}(t),\rho(t)].
\label{EQM-1}
\end{eqnarray}
Integrating (\ref{EQM-1}) we obtain
\begin{eqnarray}
\rho(t)=\rho(t_i)-i\int_{t_i}^tdt'[H_{SB}(t'),\rho(t')].
\label{EQM-2}
\end{eqnarray}
Reinserting (\ref{EQM-2}) in (\ref{EQM-1}) and tracing over the bath,
$\rho_S(t)=\text{Tr}_B\rho(t)$, we have to second order, i.e. 
the Born approximation, the equation of motion
\begin{eqnarray}
\frac{d\rho_S(t)}{dt}=
-\int_{t_i}^tdt'\text{Tr}_B[H_{SB}(t),[H_{SB}(t'),\rho(t')]].
\label{EQM-3}
\end{eqnarray}
We note that (\ref{EQM-3}) is not a proper equation of motion since
it depends on
the initial preparation at $t=t_i$. In the EQM this issue is dealt with by making
additional physical approximations. In the field theoretical approach
the Dyson equation based on a multi-oscillator bath includes secular effects
and readily yields a proper master equation of motion as shown in 
Sec. \ref{review}.

Next one makes the physical approximation 
\begin{eqnarray}
\rho(t)\approx \rho_S(t)\rho_B,
\label{EQM-4}
\end{eqnarray}
where $\rho_B$ is the static density operator for the bath. This ansatz
which provides closure is assumed valid for weak coupling 
and at all times yielding the equation of motion
\begin{eqnarray}
\frac{d\rho_S(t)}{dt}=
-\int_{t_i}^tdt'\text{Tr}_B[H_{SB}(t),[H_{SB}(t'),\rho_S(t')\rho_B]].
\label{EQM-5}
\end{eqnarray}
In order to obtain the Markov limit one replaces $\rho_S(t')$ 
in (\ref{EQM-5}) with $\rho_S(t)$, based on a time
scale separation assumption yielding the master equation
\begin{eqnarray}
\frac{d\rho_S(t)}{dt}=
-\int_{t_i}^tdt'\text{Tr}_B[H_{SB}(t),[H_{SB}(t'),\rho_S(t)\rho_B]].
\label{EQM-6}
\end{eqnarray}
Combined with additional physical approximations this 
procedure yields the Redfield equation.

In the present and standard derivations of the Redfield and Lindblad
master equations \cite{Breuer06,Tupkary22} the bath characterised
by the density operator $\rho_B$ is basically assumed to be 
unperturbed by the quantum system described by $\rho_S(t)$. The
coupling to the bath then imparts the temperature of the bath to
the quantum system. In the real world the quantum system does not 
in general relax to a Gibb's state
but will contain corrections due to the finite system-bath coupling.
These corrections can be addressed by as a starting point considering
the Gibb's state of the combined quantum system and bath.
For a detailed discussion see \cite{Thingna12}.

We note an analogy with the case of the Bohr model where
the coupling to the radiation field give rise to the Lamb shift.

\subsection{Nakajima-Zwanzig projection approach (NZ)}
The NZ  \cite{Gonzalez24}   is   based on the projection operators $P$ and $Q$ 
in order to formally separate contributions related to the system and 
to the bath, respectively. $P$ and $Q$ act in the combined Hilbert space 
for system and bath. $P$ projects out system properties, $Q$ 
refers to the bath. We have $P+Q=I$, $PP=P$, $QQ=Q$, and
$PQ=0$. In the following we also assume 
$H_{SB}=\sum_\alpha S^\alpha B^\alpha$ and the factorisation
of initial condition, $\rho(t_i)=\rho_S(t_i)\rho_B$.

Defining the  Liouville super operator $L(t)$ according to 
$L(t)\rho(t)=-i[H_{SB}(t),\rho(t)]$ the starting point is the von 
Neumann equation of motion for $\rho(t)$,
\begin{eqnarray}
\frac{d\rho(t)}{dt}=L(t)\rho(t).
\label{NZ-1}
\end{eqnarray}
Inserting $P+Q=I$ in (\ref{NZ-1}) we have
\begin{eqnarray}
&&\frac{dP\rho(t)}{dt}=PL(t)P\rho(t)+PL(t)Q\rho(t),
\label{NZ-2}
\\
&&\frac{dQ\rho(t)}{dt}=QL(t)P\rho(t)+QL(t)Q\rho(t).
\label{NZ-3}
\end{eqnarray}
Solving (\ref{NZ-3}) we find
\begin{eqnarray}
&&Q\rho(t)=G_1(t,t_i)Q\rho(0)+\int_{t_i}^t dt'G_1(t,t')QL(t')P\rho(t'),
\label{NZ-4}
\\
&&G_1(t,t_i)=\bigg(\exp\bigg[\int_{t_i}^t dt''QL(t'')\bigg]\bigg)_+,
\label{NZ-5}
\end{eqnarray}
and reinserting (\ref{NZ-4}) in (\ref{NZ-2}) we obtain the formally
exact Nakajima-Zwanzig equation for the relevant part $P\rho(t)$
\begin{eqnarray}
\frac{dP\rho(t)}{dt}=PL(t)G_1(t,t_i)Q\rho(t_i)+PL(t)P\rho(t)
+\int_{t_i}^tdt'PL(t)G_1(t,t')QL(t')P\rho(t').
\label{NZ-6}
\end{eqnarray}
The NZ equation in (\ref{NZ-6})  is formal and at this stage 
corresponds to a reformulation of the equation of motion 
(\ref{NZ-1}).

In order to proceed we specify $P$. In the NZ approach 
$P$ is given by
\begin{eqnarray}
P\rho(t)=\rho_S(t)\rho_B.
\label{NZ-7}
\end{eqnarray}
Since $\text{Tr}_B\rho_B=1$ it follows that $PP=P$.
We note that the definition of $P$ in (\ref{NZ-7})  correspond to 
the approximation $\rho(t)\approx \rho_S(t)\rho_B$ in (\ref{EQM-4}). 

Assuming that the bath is Gaussian, e.g. a multi-oscillator bath, we 
infer that $\text{Tr}_B(\rho_B B^\alpha) =0$ corresponding to 
$PL(t)P=0$. Moreover, choosing the  initial condition
$\rho(t_i)=\rho_S(t_i)\rho_B$ we have $Q\rho(t_i)=(I-P)\rho(t_i)=0$
and the NZ equation (\ref{NZ-6}) takes the simpler form
\begin{eqnarray}
&&\frac{dP\rho(t)}{dt}=\int_{t_i}^tdt'K(t,t')P\rho(t')
\label{NZ-8}
\\
&&K(t,t')=PL(t)G_1(t,t')QL(t')P
\label{NZ-9}
\end{eqnarray}
To leading order $G_1(t,t')=I$ and we obtain, using $PL(t)P=0$,
the kernel $K(t,t')=PL(t)L(t')P$, i.e. the equation of motion
\begin{eqnarray}
\frac{dP\rho(t)}{dt}=\int_{t_i}^tdt'PL(t)L(t')P\rho(t').
\label{NZ-10}
\end{eqnarray}
Inserting the definition of $L$ we obtain the equation of motion
(\ref{EQM-5}) and we infer that the NZ method to second order is 
equivalent to the EQM method. 
\subsection{Time-convolutionsless projection method (TCL)}
The TCL is an extension of the NZ which yields a perturbation
scheme directly giving rise to a master equation in the Markov limit
\cite{Ferguson21,Gu24}. 
In the second order equation of motion  (\ref{NZ-10}) $\rho_S(t')$ 
samples
the past history from the initial time $t_i$ to $t$. In the TCL 
approach the procedure is to
propagate $\rho_S(t')$ forward to the final time $t$ in order to attain
the Markov limit directly.

Using $Q\rho(t_i)=0$ we have from (\ref{NZ-4})
\begin{eqnarray}
Q\rho(t)=\int_{t_i}^t dt'G_1(t,t')QL(t')P\rho(t').
\label{TCL-1}
\end{eqnarray}
In order to propagate $P\rho(t')$ to $P\rho(t)$, where $t'<t$
we invert the solution of (\ref{NZ-1}), i.e.
\begin{eqnarray}
&&\rho(t')=G_2(t,t')(P+Q)\rho(t),
\label{TCL-2}
\\
&&G_2(t,t')=\bigg(\exp\bigg[-\int_{t'}^t dt''QL(t'')\bigg]\bigg)_-,
\label{TCL-3}
\end{eqnarray}
where $(\cdots)_-$ indicates the anti-chronological time order.
Inserting $\rho(t')$ in (\ref{TCL-1}), solving for $Q\rho(t)$ and inserting
in (\ref{NZ-8}) we obtain the Markov master equation
\begin{eqnarray}
&&\frac{dP\rho(t)}{dt}=K(t)P\rho(t),
\label{TCL-4}
\\
&&K(t)=PL(t)(1-\Sigma(t))^{-1}P,
\label{TCL-5}
\\
&&\Sigma(t)=\int_{t_i}^tdt'G_1(t,t')QL(t')PG_2(t,t'),
\label{TCL-6}
\end{eqnarray}
Here $K(t)$ is the so-called TCL generator.

Expanding $(1-\Sigma(t))^{-1}$ the TCL scheme allows for a systematic 
perturbation theory in powers of the system-bath interaction. To leading
order we have
\begin{eqnarray}
K(t)=\int_{t_i}^tdt' PL(t)L(t')P
\label{TCL-7}
\end{eqnarray}
and inserting $L(t)$ the master equation in (\ref{EQM-6}).
For further details we refer to \cite{Breuer06}. 
\newpage
\begin{figure}
\label{fig}
\begin{center}
\includegraphics[width=1.0\textwidth]{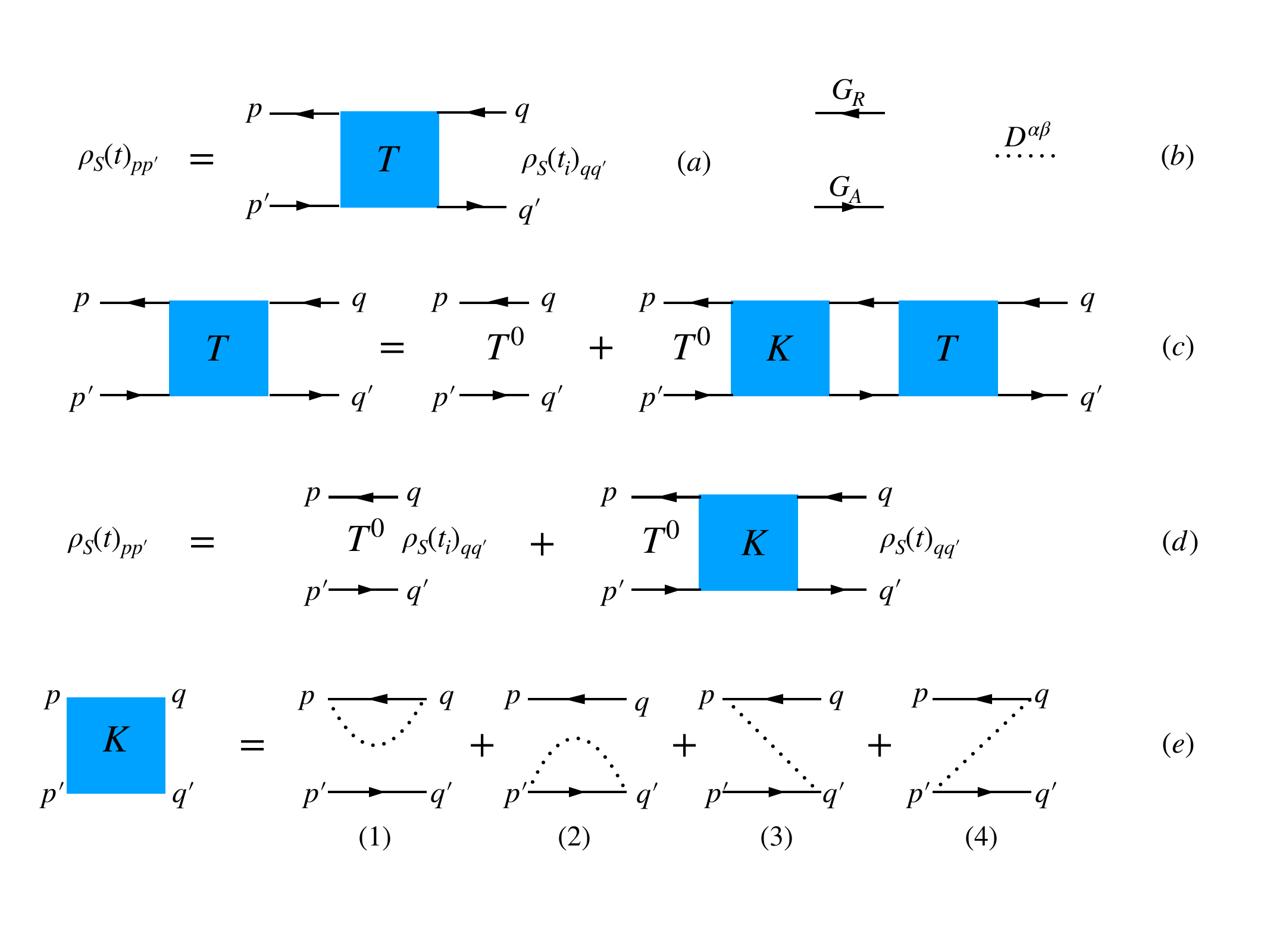}
\end{center}
\caption{In Fig. a we depict the transmission matrix $T(t,t_i)_{pp',qq'}$ in 
(\ref{density}) showing the evolution of the density matrix 
$\rho_S(t)_{pp'}$ from the initial time $t=t_i$ to the final time $t$.
In Fig. b we show the retarded and advanced Green's functions
$G_R(t,t')_{pq}$ and $G_A(t,t')_{pq}$ and the bath correlation
function $D^{\alpha\beta}(t,t')$ in (\ref{greenret}), (\ref{greenadv}),
and (\ref{bath}), respectively. In Fig. c we depict the Dyson equation
in (\ref{dyson}) satisfied by the transmission matrix $T$. Here $T^0$ 
is unperturbed transmission matrix in (\ref{free}) and $K$ the
irreversible kernel. In Fig. d we show the inhomogeneous integral
equation for $\rho_S(t)_{pp'}$ in (\ref{intmaster}) with 
inhomogeneous term $T^0\rho_S(t_i)$ and integral $T^0K\rho_S(t_i)$.
In Fig. e we depict in (\ref{kernelborn}) the second order Born
contribution to the irreversible kernel $K$. The terms (1) and (2) 
correspond to a renormalisation of the retarded and advanced
Green's functions, respectively. The terms  (3) and (4) correspond to
cross correlations.}
\end{figure}
\clearpage

\begin{thebibliography}{44}
\expandafter\ifx\csname natexlab\endcsname\relax\def\natexlab#1{#1}\fi
\expandafter\ifx\csname bibnamefont\endcsname\relax
  \def\bibnamefont#1{#1}\fi
\expandafter\ifx\csname bibfnamefont\endcsname\relax
  \def\bibfnamefont#1{#1}\fi
\expandafter\ifx\csname citenamefont\endcsname\relax
  \def\citenamefont#1{#1}\fi
\expandafter\ifx\csname url\endcsname\relax
  \def\url#1{\texttt{#1}}\fi
\expandafter\ifx\csname urlprefix\endcsname\relax\def\urlprefix{URL }\fi
\providecommand{\bibinfo}[2]{#2}
\providecommand{\eprint}[2][]{\url{#2}}

\bibitem[{\citenamefont{Horodecki et~al.}(2009)\citenamefont{Horodecki,
  Horodecki, Horodecki, and Horodecki}}]{Horodecki09}
\bibinfo{author}{\bibfnamefont{R.}~\bibnamefont{Horodecki}},
  \bibinfo{author}{\bibfnamefont{P.}~\bibnamefont{Horodecki}},
  \bibinfo{author}{\bibfnamefont{M.}~\bibnamefont{Horodecki}},
  \bibnamefont{and}
  \bibinfo{author}{\bibfnamefont{K.}~\bibnamefont{Horodecki}},
  \bibinfo{journal}{Rev. Mod. Phys.} \textbf{\bibinfo{volume}{81}},
  \bibinfo{pages}{865} (\bibinfo{year}{2009}).

\bibitem[{\citenamefont{Paneru et~al.}(2020)\citenamefont{Paneru, Cohen,
  Fickler, Boyd, and Karimi}}]{Paneru20}
\bibinfo{author}{\bibfnamefont{D.}~\bibnamefont{Paneru}},
  \bibinfo{author}{\bibfnamefont{E.}~\bibnamefont{Cohen}},
  \bibinfo{author}{\bibfnamefont{R.}~\bibnamefont{Fickler}},
  \bibinfo{author}{\bibfnamefont{R.~W.} \bibnamefont{Boyd}}, \bibnamefont{and}
  \bibinfo{author}{\bibfnamefont{E.}~\bibnamefont{Karimi}},
  \bibinfo{journal}{Rep. Prog. Phys.} \textbf{\bibinfo{volume}{83}},
  \bibinfo{pages}{064001} (\bibinfo{year}{2020}).

\bibitem[{\citenamefont{Bresque et~al.}(2021)\citenamefont{Bresque, Camati,
  Rogers, Murch, Jordan, and Auffeves}}]{Bresque21}
\bibinfo{author}{\bibfnamefont{L.}~\bibnamefont{Bresque}},
  \bibinfo{author}{\bibfnamefont{P.~A.} \bibnamefont{Camati}},
  \bibinfo{author}{\bibfnamefont{S.}~\bibnamefont{Rogers}},
  \bibinfo{author}{\bibfnamefont{K.}~\bibnamefont{Murch}},
  \bibinfo{author}{\bibfnamefont{A.~N.} \bibnamefont{Jordan}},
  \bibnamefont{and} \bibinfo{author}{\bibfnamefont{A.}~\bibnamefont{Auffeves}},
  \bibinfo{journal}{Phys. Rev. Lett.} \textbf{\bibinfo{volume}{126}},
  \bibinfo{pages}{120605} (\bibinfo{year}{2021}).

\bibitem[{\citenamefont{Vinjanampathy and Anders}(2016)}]{Vinjanampathy16}
\bibinfo{author}{\bibfnamefont{S.}~\bibnamefont{Vinjanampathy}}
  \bibnamefont{and} \bibinfo{author}{\bibfnamefont{J.}~\bibnamefont{Anders}},
  \bibinfo{journal}{Contemporary Physics} \textbf{\bibinfo{volume}{57}},
  \bibinfo{pages}{545} (\bibinfo{year}{2016}).

\bibitem[{\citenamefont{Kosloff and Levy}(2019)}]{Kosloff19}
\bibinfo{author}{\bibfnamefont{R.}~\bibnamefont{Kosloff}} \bibnamefont{and}
  \bibinfo{author}{\bibfnamefont{A.}~\bibnamefont{Levy}}, \bibinfo{journal}{J.
  Chem. Phys.} \textbf{\bibinfo{volume}{150}}, \bibinfo{pages}{204105}
  (\bibinfo{year}{2019}).

\bibitem[{\citenamefont{Rivas}(2020)}]{Rivas20}
\bibinfo{author}{\bibfnamefont{A.}~\bibnamefont{Rivas}},
  \bibinfo{journal}{Phys. Rev. Lett.} \textbf{\bibinfo{volume}{124}},
  \bibinfo{pages}{160601} (\bibinfo{year}{2020}).

\bibitem[{\citenamefont{Breuer and Petruccione}(2006)}]{Breuer06}
\bibinfo{author}{\bibfnamefont{H.~P.} \bibnamefont{Breuer}} \bibnamefont{and}
  \bibinfo{author}{\bibfnamefont{F.}~\bibnamefont{Petruccione}},
  \emph{\bibinfo{title}{The Theory of Open Quantum Systems}}
  (\bibinfo{publisher}{Oxford University Press}, \bibinfo{address}{Oxford},
  \bibinfo{year}{2006}).

\bibitem[{\citenamefont{Redfield}(1965)}]{Redfield65}
\bibinfo{author}{\bibfnamefont{A.~G.} \bibnamefont{Redfield}},
  \bibinfo{journal}{Advances in Magnetic and Optical Resonance}
  \textbf{\bibinfo{volume}{1}}, \bibinfo{pages}{1} (\bibinfo{year}{1965}).

\bibitem[{\citenamefont{Gonzalez-Ballestero}(2024)}]{Gonzalez24}
\bibinfo{author}{\bibfnamefont{C.}~\bibnamefont{Gonzalez-Ballestero}},
  \bibinfo{journal}{Quantum 8} \textbf{\bibinfo{volume}{49}},
  \bibinfo{pages}{1454} (\bibinfo{year}{2024}).

\bibitem[{\citenamefont{Breuer and Kappler}(2001)}]{Breuer01}
\bibinfo{author}{\bibfnamefont{H.~P.} \bibnamefont{Breuer}} \bibnamefont{and}
  \bibinfo{author}{\bibfnamefont{B.}~\bibnamefont{Kappler}},
  \bibinfo{journal}{Ann. Phys. (NY)} \textbf{\bibinfo{volume}{291}},
  \bibinfo{pages}{36} (\bibinfo{year}{2001}).

\bibitem[{\citenamefont{Lindblad}(1976)}]{Lindblad76}
\bibinfo{author}{\bibfnamefont{G.}~\bibnamefont{Lindblad}},
  \bibinfo{journal}{Commun. Math. Phys.} \textbf{\bibinfo{volume}{48}},
  \bibinfo{pages}{119} (\bibinfo{year}{1976}).

\bibitem[{\citenamefont{Manzano}(2020)}]{Manzano20}
\bibinfo{author}{\bibfnamefont{D.}~\bibnamefont{Manzano}},
  \bibinfo{journal}{AIP Advances} \textbf{\bibinfo{volume}{10}},
  \bibinfo{pages}{025106} (\bibinfo{year}{2020}).

\bibitem[{\citenamefont{Chruscinski and Pascazio}(2017)}]{Chrus17}
\bibinfo{author}{\bibfnamefont{D.}~\bibnamefont{Chruscinski}} \bibnamefont{and}
  \bibinfo{author}{\bibfnamefont{S.}~\bibnamefont{Pascazio}},
  \bibinfo{journal}{Open Systems and Information Dynamics}
  \textbf{\bibinfo{volume}{24}}, \bibinfo{pages}{1740001}
  (\bibinfo{year}{2017}).

\bibitem[{\citenamefont{Tupkary et~al.}(2022)\citenamefont{Tupkary, Dhar,
  Kulkarni, and Purkayastha}}]{Tupkary22}
\bibinfo{author}{\bibfnamefont{D.}~\bibnamefont{Tupkary}},
  \bibinfo{author}{\bibfnamefont{A.}~\bibnamefont{Dhar}},
  \bibinfo{author}{\bibfnamefont{M.}~\bibnamefont{Kulkarni}}, \bibnamefont{and}
  \bibinfo{author}{\bibfnamefont{A.}~\bibnamefont{Purkayastha}},
  \bibinfo{journal}{Phys. Rev. A} \textbf{\bibinfo{volume}{105}},
  \bibinfo{pages}{032208} (\bibinfo{year}{2022}).

\bibitem[{\citenamefont{Gorini et~al.}(1976)\citenamefont{Gorini, Kossakowski,
  and Sudarshan}}]{Gorini76}
\bibinfo{author}{\bibfnamefont{V.}~\bibnamefont{Gorini}},
  \bibinfo{author}{\bibfnamefont{A.}~\bibnamefont{Kossakowski}},
  \bibnamefont{and} \bibinfo{author}{\bibfnamefont{E.~C.~G.}
  \bibnamefont{Sudarshan}}, \bibinfo{journal}{J. Math. Phys.}
  \textbf{\bibinfo{volume}{17}}, \bibinfo{pages}{821} (\bibinfo{year}{1976}).

\bibitem[{\citenamefont{Agredo et~al.}(2021)\citenamefont{Agredo, Fagnola, and
  Poletti}}]{Kossakowski21}
\bibinfo{author}{\bibfnamefont{J.}~\bibnamefont{Agredo}},
  \bibinfo{author}{\bibfnamefont{F.}~\bibnamefont{Fagnola}}, \bibnamefont{and}
  \bibinfo{author}{\bibfnamefont{D.}~\bibnamefont{Poletti}},
  \bibinfo{journal}{Open Systems and Information Dynamics}
  \textbf{\bibinfo{volume}{28}}, \bibinfo{pages}{2170001}
  (\bibinfo{year}{2021}).

\bibitem[{\citenamefont{Sudarshan}(1963)}]{Sudarshan63}
\bibinfo{author}{\bibfnamefont{E.~C.~G.} \bibnamefont{Sudarshan}},
  \bibinfo{journal}{Phys. Rev. Lett.} \textbf{\bibinfo{volume}{10}},
  \bibinfo{pages}{277} (\bibinfo{year}{1963}).

\bibitem[{\citenamefont{Davies}(1974)}]{Davies74}
\bibinfo{author}{\bibfnamefont{E.~B.} \bibnamefont{Davies}},
  \bibinfo{journal}{Commun. Math. Phys.} \textbf{\bibinfo{volume}{39}},
  \bibinfo{pages}{91} (\bibinfo{year}{1974}).

\bibitem[{\citenamefont{Fogedby}(2022)}]{Fogedby22}
\bibinfo{author}{\bibfnamefont{H.~C.} \bibnamefont{Fogedby}},
  \bibinfo{journal}{Phys. Rev. A} \textbf{\bibinfo{volume}{106}},
  \bibinfo{pages}{022205} (\bibinfo{year}{2022}).

\bibitem[{\citenamefont{Caldeira and
  Leggett}(1983{\natexlab{a}})}]{Caldeira83a}
\bibinfo{author}{\bibfnamefont{A.~O.} \bibnamefont{Caldeira}} \bibnamefont{and}
  \bibinfo{author}{\bibfnamefont{A.~J.} \bibnamefont{Leggett}},
  \bibinfo{journal}{Ann. Phys.} \textbf{\bibinfo{volume}{149}},
  \bibinfo{pages}{374} (\bibinfo{year}{1983}{\natexlab{a}}).

\bibitem[{\citenamefont{Zhang et~al.}(2024)\citenamefont{Zhang, Denker,
  Asadian, and hne}}]{Zhang24}
\bibinfo{author}{\bibfnamefont{C.}~\bibnamefont{Zhang}},
  \bibinfo{author}{\bibfnamefont{S.}~\bibnamefont{Denker}},
  \bibinfo{author}{\bibfnamefont{A.}~\bibnamefont{Asadian}}, \bibnamefont{and}
  \bibinfo{author}{\bibfnamefont{O.~G.} \bibnamefont{hne}},
  \bibinfo{journal}{Phys. Rev. Lett.} \textbf{\bibinfo{volume}{133}},
  \bibinfo{pages}{040203} (\bibinfo{year}{2024}).

\bibitem[{\citenamefont{Gaudin}(1960)}]{Gaudin60}
\bibinfo{author}{\bibfnamefont{M.}~\bibnamefont{Gaudin}},
  \bibinfo{journal}{Nuclear Physics} \textbf{\bibinfo{volume}{15}},
  \bibinfo{pages}{89} (\bibinfo{year}{1960}).

\bibitem[{\citenamefont{Mahan}(1990)}]{Mahan90}
\bibinfo{author}{\bibfnamefont{G.~D.} \bibnamefont{Mahan}},
  \emph{\bibinfo{title}{Many Particle Physics}} (\bibinfo{publisher}{Plenum
  Press}, \bibinfo{address}{New York}, \bibinfo{year}{1990}).

\bibitem[{\citenamefont{Purkayastha and M{\o}lmer}(2023)}]{Purkayastha23}
\bibinfo{author}{\bibfnamefont{A.}~\bibnamefont{Purkayastha}} \bibnamefont{and}
  \bibinfo{author}{\bibfnamefont{K.}~\bibnamefont{M{\o}lmer}},
  \bibinfo{journal}{Phys. Rev. A} \textbf{\bibinfo{volume}{108}},
  \bibinfo{pages}{053704} (\bibinfo{year}{2023}).

\bibitem[{\citenamefont{Clerk and Sipe}(1998)}]{Clerk98}
\bibinfo{author}{\bibfnamefont{A.~A.} \bibnamefont{Clerk}} \bibnamefont{and}
  \bibinfo{author}{\bibfnamefont{J.~E.} \bibnamefont{Sipe}},
  \bibinfo{journal}{Found. Phys.} \textbf{\bibinfo{volume}{28}},
  \bibinfo{pages}{639} (\bibinfo{year}{1998}).

\bibitem[{\citenamefont{J{\o}rgensen and Wubs}(2022)}]{Jorgensen22}
\bibinfo{author}{\bibfnamefont{M.~A.} \bibnamefont{J{\o}rgensen}}
  \bibnamefont{and} \bibinfo{author}{\bibfnamefont{M.}~\bibnamefont{Wubs}},
  \bibinfo{journal}{J. Phys. B: At. Mol. Opt. Phys.}
  \textbf{\bibinfo{volume}{55}}, \bibinfo{pages}{195401}
  (\bibinfo{year}{2022}).

\bibitem[{\citenamefont{Nathan and Rudner}(2020)}]{Nathan20}
\bibinfo{author}{\bibfnamefont{F.}~\bibnamefont{Nathan}} \bibnamefont{and}
  \bibinfo{author}{\bibfnamefont{M.~S.} \bibnamefont{Rudner}},
  \bibinfo{journal}{Phys. Rev. B} \textbf{\bibinfo{volume}{102}},
  \bibinfo{pages}{115109} (\bibinfo{year}{2020}).

\bibitem[{\citenamefont{Nathan and Rudner}(2021)}]{Nathan21}
\bibinfo{author}{\bibfnamefont{F.}~\bibnamefont{Nathan}} \bibnamefont{and}
  \bibinfo{author}{\bibfnamefont{M.~S.} \bibnamefont{Rudner}},
  \bibinfo{journal}{Erratum Phys. Rev. B} \textbf{\bibinfo{volume}{104}},
  \bibinfo{pages}{119901} (\bibinfo{year}{2021}).

\bibitem[{\citenamefont{Nathan}(2018)}]{Nathan18}
\bibinfo{author}{\bibfnamefont{F.}~\bibnamefont{Nathan}}
  (\bibinfo{year}{2018}), \bibinfo{note}{phD Thesis, University of Copenhagen}.

\bibitem[{\citenamefont{Lee and Yeo}(2020)}]{Lee20}
\bibinfo{author}{\bibfnamefont{J.~S.} \bibnamefont{Lee}} \bibnamefont{and}
  \bibinfo{author}{\bibfnamefont{J.}~\bibnamefont{Yeo}} (\bibinfo{year}{2020}),
  \bibinfo{note}{arXiv:2011.00735}.

\bibitem[{\citenamefont{Nathan and Rudner}(2024)}]{Nathan22}
\bibinfo{author}{\bibfnamefont{F.}~\bibnamefont{Nathan}} \bibnamefont{and}
  \bibinfo{author}{\bibfnamefont{M.~S.} \bibnamefont{Rudner}},
  \bibinfo{journal}{Phys. Rev. B} \textbf{\bibinfo{volume}{109}},
  \bibinfo{pages}{205140} (\bibinfo{year}{2024}).

\bibitem[{\citenamefont{Kirsanskas et~al.}(2018)\citenamefont{Kirsanskas,
  Franckie, and Wacker}}]{Kirsanskas18}
\bibinfo{author}{\bibfnamefont{G.}~\bibnamefont{Kirsanskas}},
  \bibinfo{author}{\bibfnamefont{M.}~\bibnamefont{Franckie}}, \bibnamefont{and}
  \bibinfo{author}{\bibfnamefont{A.}~\bibnamefont{Wacker}},
  \bibinfo{journal}{Phys. Rev. B} \textbf{\bibinfo{volume}{97}},
  \bibinfo{pages}{035432} (\bibinfo{year}{2018}).

\bibitem[{\citenamefont{Nathan et~al.}(2019)\citenamefont{Nathan, Martin, and
  Refael}}]{Nathan19}
\bibinfo{author}{\bibfnamefont{F.}~\bibnamefont{Nathan}},
  \bibinfo{author}{\bibfnamefont{I.}~\bibnamefont{Martin}}, \bibnamefont{and}
  \bibinfo{author}{\bibfnamefont{G.}~\bibnamefont{Refael}},
  \bibinfo{journal}{Phys. Rev. B} \textbf{\bibinfo{volume}{99}},
  \bibinfo{pages}{094311} (\bibinfo{year}{2019}).

\bibitem[{\citenamefont{Davidovich}(2020)}]{Davidovich20}
\bibinfo{author}{\bibfnamefont{D.}~\bibnamefont{Davidovich}},
  \bibinfo{journal}{Quantum} \textbf{\bibinfo{volume}{4}}, \bibinfo{pages}{326}
  (\bibinfo{year}{2020}).

\bibitem[{\citenamefont{McDonald and Clerk}(2023)}]{McDonald23}
\bibinfo{author}{\bibfnamefont{A.}~\bibnamefont{McDonald}} \bibnamefont{and}
  \bibinfo{author}{\bibfnamefont{A.~A.} \bibnamefont{Clerk}},
  \bibinfo{journal}{Phys. Rev. Research} \textbf{\bibinfo{volume}{5}},
  \bibinfo{pages}{033107} (\bibinfo{year}{2023}).

\bibitem[{\citenamefont{Reimer and Wegewijs}(2019)}]{Reimer19a}
\bibinfo{author}{\bibfnamefont{V.}~\bibnamefont{Reimer}} \bibnamefont{and}
  \bibinfo{author}{\bibfnamefont{M.~R.} \bibnamefont{Wegewijs}},
  \bibinfo{journal}{SciPost Phys.} \textbf{\bibinfo{volume}{7}},
  \bibinfo{pages}{1} (\bibinfo{year}{2019}).

\bibitem[{\citenamefont{Thompson and Kamenev}(2023)}]{Thompson23}
\bibinfo{author}{\bibfnamefont{F.}~\bibnamefont{Thompson}} \bibnamefont{and}
  \bibinfo{author}{\bibfnamefont{A.}~\bibnamefont{Kamenev}},
  \bibinfo{journal}{Annals of Physics} \textbf{\bibinfo{volume}{455}},
  \bibinfo{pages}{169385} (\bibinfo{year}{2023}).

\bibitem[{\citenamefont{Sieberer et~al.}(2016)\citenamefont{Sieberer, Buchhold,
  and Diehl}}]{Sieberer16}
\bibinfo{author}{\bibfnamefont{L.~M.} \bibnamefont{Sieberer}},
  \bibinfo{author}{\bibfnamefont{M.}~\bibnamefont{Buchhold}}, \bibnamefont{and}
  \bibinfo{author}{\bibfnamefont{S.}~\bibnamefont{Diehl}},
  \bibinfo{journal}{Rep. Prog. Phys.} \textbf{\bibinfo{volume}{79}},
  \bibinfo{pages}{096001} (\bibinfo{year}{2016}).

\bibitem[{\citenamefont{Gu}(2024)}]{Gu24}
\bibinfo{author}{\bibfnamefont{B.}~\bibnamefont{Gu}}, \bibinfo{journal}{J.
  Chem. Phys.} \textbf{\bibinfo{volume}{160}}, \bibinfo{pages}{204113}
  (\bibinfo{year}{2024}).

\bibitem[{\citenamefont{Ferguson et~al.}(2021)\citenamefont{Ferguson,
  Zilberberg, and Blatter}}]{Ferguson21}
\bibinfo{author}{\bibfnamefont{M.~S.} \bibnamefont{Ferguson}},
  \bibinfo{author}{\bibfnamefont{O.}~\bibnamefont{Zilberberg}},
  \bibnamefont{and} \bibinfo{author}{\bibfnamefont{G.}~\bibnamefont{Blatter}},
  \bibinfo{journal}{Phys. Rev. Research} \textbf{\bibinfo{volume}{3}},
  \bibinfo{pages}{023127} (\bibinfo{year}{2021}).

\bibitem[{\citenamefont{Nestmann and Wegewijs}(2021)}]{Nestmann21a}
\bibinfo{author}{\bibfnamefont{K.}~\bibnamefont{Nestmann}} \bibnamefont{and}
  \bibinfo{author}{\bibfnamefont{M.~R.} \bibnamefont{Wegewijs}},
  \bibinfo{journal}{Phys. Rev. B} \textbf{\bibinfo{volume}{104}},
  \bibinfo{pages}{155407} (\bibinfo{year}{2021}).

\bibitem[{\citenamefont{Nestmann et~al.}(2021)\citenamefont{Nestmann, Bruch,
  and Wegewijs}}]{Nestmann21b}
\bibinfo{author}{\bibfnamefont{K.}~\bibnamefont{Nestmann}},
  \bibinfo{author}{\bibfnamefont{V.}~\bibnamefont{Bruch}}, \bibnamefont{and}
  \bibinfo{author}{\bibfnamefont{M.~R.} \bibnamefont{Wegewijs}},
  \bibinfo{journal}{Phys. Rev. X} \textbf{\bibinfo{volume}{11}},
  \bibinfo{pages}{021041} (\bibinfo{year}{2021}).

\bibitem[{\citenamefont{Caldeira and
  Leggett}(1983{\natexlab{b}})}]{Caldeira83b}
\bibinfo{author}{\bibfnamefont{A.~O.} \bibnamefont{Caldeira}} \bibnamefont{and}
  \bibinfo{author}{\bibfnamefont{A.~J.} \bibnamefont{Leggett}},
  \bibinfo{journal}{Physica} \textbf{\bibinfo{volume}{121A}},
  \bibinfo{pages}{587} (\bibinfo{year}{1983}{\natexlab{b}}).

\bibitem[{\citenamefont{Thingna et~al.}(2012)\citenamefont{Thingna, Wang, and
  H\"{a}nggi}}]{Thingna12}
\bibinfo{author}{\bibfnamefont{J.}~\bibnamefont{Thingna}},
  \bibinfo{author}{\bibfnamefont{J.-S.} \bibnamefont{Wang}}, \bibnamefont{and}
  \bibinfo{author}{\bibfnamefont{P.}~\bibnamefont{H\"{a}nggi}},
  \bibinfo{journal}{J. Chem. Phys.} \textbf{\bibinfo{volume}{136}},
  \bibinfo{pages}{1587} (\bibinfo{year}{2012}).

\end{thebibliography}

\end{document}